\documentclass[%
  twoside,
  reprint,
  amsmath,amssymb,
  aps,
  pra,
  nofootinbib,
  showpacs,
  a4paper
]{revtex4-1}

\usepackage{graphicx}%
\usepackage[usenames,dvipsnames]{xcolor}
\usepackage{subfigure}
\usepackage{siunitx}
\DeclareSIUnit{\au}{{a.u.}}

\usepackage{enumitem} 
\usepackage{rotating}
\usepackage{empheq} 
\usepackage{multirow} 

\usepackage[
text={7.3in,10in},centering,
total={6.5in,8.75in}, top=1in, left=0.52in, includefoot,
]{geometry}

\usepackage[
  bookmarks=true,
  colorlinks,
  linkcolor=blue,
  urlcolor=blue,
  citecolor=blue,
  plainpages=false,
  pdfpagelabels,
  final,
  breaklinks=true
]{hyperref}
\hypersetup{
pdftitle={Slalom in complex time: emergence of low-energy structures in tunnel ionization via complex time contours}, 
pdfauthor={Emilio Pisanty and Misha Ivanov}
}

\usepackage{natbib}
\makeatletter \def\NAT@def@citea{\def\@citea{\NAT@separator\,}} \makeatother
\newcommand{\citer}[1]{Ref.~\citealp{#1}}
\newcommand{\citers}[1]{Refs.~\citealp{#1}}

\newcommand{\reffig}[1]{Fig.~\ref{#1}}

\newcommand{\figureonegamma}{0.75}

\newcommand{\onemicronparsfield}{0.05} 
\newcommand{\onemicronparswavelength}{1} 
\newcommand{\onemicronparskappa}{1.07} 
\newcommand{\onemicronparsgamma}{0.98}

\newcommand{\figurefivepo}{0.02} 
\newcommand{\figurefivepp}{0.8}
\newcommand{\figuretwelvepo}{0.001}
\newcommand{\figuretwelveppl}{0.063}
\newcommand{\figuretwelvepph}{0.0635}
\newcommand{\figurethirteenapo}{0.1}
\newcommand{\figurethirteenapp}{0.2}
\newcommand{\figurethirteenbpo}{0.02}
\newcommand{\figurethirteenbpp}{0.8}
\newcommand{\figurethirteencpo}{0.001}
\newcommand{\figurethirteencpp}{0.0635 F/\omega}
\newcommand{\figurethirteendpo}{0.05}
\newcommand{\figurethirteendpp}{1.1}

\newcommand{\figurefourteenfirstpztransition}{0.063} 
\newcommand{\figurefourteenthirdpztransition}{0.031} 
 
\newcommand{\figurefourteenwavelength}{3.1} 
 
\newcommand{\figurefourteengamma}{0.31}
\newcommand{\figurefifteenpomult}{1.07}

\renewcommand{\d}{\ensuremath{\textrm{d}}}
\renewcommand{\Re}{\operatorname{Re}}
\renewcommand{\Im}{\operatorname{Im}}
\newcommand{\bra}[1]{\ensuremath{\left\langle#1\right|}}
\newcommand{\ket}[1]{\ensuremath{\left|#1\right\rangle}}
\newcommand{\bracket}[2]{\ensuremath{\left\langle#1 \vphantom{#2}\middle|  #2 \vphantom{#1}\right\rangle}}
\newcommand{\matrixel}[3]{\ensuremath{\left\langle #1 \vphantom{#2#3} \middle| #2 \middle| #3 \vphantom{#1#2} \right\rangle}}
\newcommand{\vb}[1]{\mathbf{#1}}
  \newcommand{\vbv}{\vb{v}}
  \newcommand{\vbr}{\vb{r}}
  \newcommand{\vba}{\vb{A}}
  \newcommand{\vbf}{\vb{F}}
  \newcommand{\vbp}{\vb{p}}

\newcommand{\uz}{\hat{\vb{z}}}
\newcommand{\cl}{{\mathrm{cl}}}
\newcommand{\rcl}{\vbr_\cl}
\newcommand{\rl}{\vbr_\mathrm{L}}
\newcommand{\peva}{\vbp^{\scriptscriptstyle \mathrm{EVA}}}
\newcommand{\zcl}{z_\cl}
\newcommand{\zexit}{z_\mathrm{exit}}
\newcommand{\zquiv}{z_\mathrm{quiv}}

\newcommand{\odd}{odd-$n${}}
\newcommand{\even}{even-$n${}}

\newcommand{\ts}{{t_s}}
\newcommand{\tn}{{t_0}}
\newcommand{\tauT}{{\tau_\mathrm{T}}}
\newcommand{\tr}{{t_r}}
\newcommand{\tk}{{t_\kappa}}
\newcommand{\tca}{{t_{\scriptscriptstyle\mathrm{CA}}}}
\newcommand{\tcasup}[1]{{t_{\scriptscriptstyle\mathrm{CA}}^{{#1}}}}
\newcommand{\tcacl}{{t_{\scriptscriptstyle\mathrm{CA}}^{\scriptscriptstyle\mathrm{clas}}}}
\newcommand{\pzsr}{p_z^{\mathrm{sr}}}

\DeclareMathOperator{\arcsinh}{arcsinh}

\begin{document}

\title{Slalom in complex time: emergence of low-energy structures in tunnel ionization via complex time contours}

\author{Emilio Pisanty$^{1}$}
 \email{e.pisanty11@imperial.ac.uk}
\author{Misha Ivanov$^{1,2,3}$}%
 \email{m.ivanov@imperial.ac.uk}
\affiliation{%
\scriptsize
$^1$Blackett Laboratory, Imperial College London, South Kensington Campus, SW7 2AZ London, United Kingdom\\
$^2$Department of Physics, Humboldt University, Newtonstrasse 15, 12489 Berlin, Germany\\
$^3$Max Born Institute, Max Born Strasse 2a, 12489 Berlin, Germany
}

\date{\today}

\begin{abstract}
The ionization of atoms by strong, low-frequency fields can generally be described well by assuming that the photoelectron is, after the ionization step, completely at the mercy of the laser field. However, certain phenomena, like the recent discovery of low-energy structures in the long-wavelength regime, require the inclusion of the Coulomb interaction with the ion once the electron is in the continuum. We explore the first-principles inclusion of this interaction, known as analytical $R$-matrix theory, and its consequences on the corresponding quantum orbits. We show that the trajectory must have an imaginary component, and that this causes branch cuts in the complex time plane when the real trajectory revisits the neighbourhood of the ionic core. We provide a framework for consistently navigating these branch cuts based on closest-approach times, which satisfy the equation $\mathbf{r}(t)\cdot\mathbf{v}(t)=0$ in the complex plane. We explore the geometry of these roots and describe the geometrical structures underlying the emergence of LES in both the classical and quantum domains.
\end{abstract}
\maketitle

The interaction of atoms and molecules with intense lasers is a rich field, and provides challenges for theory to describe non-perturbative phenomena which often bridge a large range of energy scales. A recent example of this is the discovery of low-energy structures (LES) \cite{blaga_original_LES,VLES_initial} in photoionization by strong, long-wavelength fields where, in addition to above-threshold electrons that absorb many more photons than required to reach the continuum, a strong peak is observed at energies far below the mean oscillation energy of electrons in the field.

This was unexpected \cite{faisal_ionization_surprise}, as the strong-field approximation (SFA) \cite{keldysh_ionization_1965} typically predicts a smooth, featureless spectrum at low energies, and SFA is generally expected to improve in accuracy as the frequency decreases and the system goes deeper into the optical tunnelling regime. This triggered active interest in developing theoretical methods to describe the LES, and in identifying the physical mechanisms that create them.

Numerical simulations of the time-dependent Schr\"o\-din\-ger equation (TDSE) do reproduce the structure \cite{blaga_original_LES}, though they are particularly demanding at long wavelengths. Simulations done with and without the ion's long-range Coulomb potential \cite{telnov_TDSE_with_and_without_Coulomb, VLES_characterization} indicate that its role is essential in producing the LES. Similarly, Monte Carlo simulations \cite{VLES_characterization,CTMC1,CTMC2,CTMC3} using classical trajectories involving both the laser and the Coulomb field support this conclusion.

From a semiclassical perspective, it is indeed possible to include the effect of the ion's potential when the electron is already in the continuum. This can be done via a perturbative Born series, like those used to explain high-order above-threshold ionization \cite{improvedSFA}; here the LES emerges as electrons that forward-scatter once at the ion \cite{Milosevic_reexamination,Milosevic_scattering_large,LES_Scaling,Milosevic_LFA}. Alternatively, the SFA can be reformulated to include the effect of the Coulomb field on the underlying trajectories~\cite{CCSFA_initial_short, CCSFA_initial_full,TCSFA_sub_barrier}; the resulting Coulomb-corrected SFA (CCSFA) also points to forward-scattered electrons \cite{yan_TCSFA_caustics}. First-principles analytical methods to include the Coulomb field's effect on the wavefunction's phase, known as analytical $R$-matrix theory (ARM) \cite{ARM_initial,ARM_circular, ARM_abinitio_verification, ARM_trajectories,ARM_attoclock,MResReport}, are as yet untested in this regime.

Analysis of the classical trajectories involved in the scattering points more specifically at \textit{soft} recollisions, where the electron does not hit the core head-on, but is instead softly deflected as it approaches the ion with nearly zero velocity, near a turning point of the laser-driven trajectory \cite{Rost_PRL,Rost_JPhysB}. Although these trajectories spend enough time near the ion that the effect of its potential is no longer perturbative (and can in fact cause chaotic dynamics \cite{chaotic_dynamics}), a comparison of the different models suggests that the LES are rooted in the pure laser-driven orbits of the simple-man's model, and that the role of the Coulomb potential is to enhance their contribution \cite{Becker_rescattering}.

\begin{figure}[h]
  \includegraphics[width=\columnwidth]{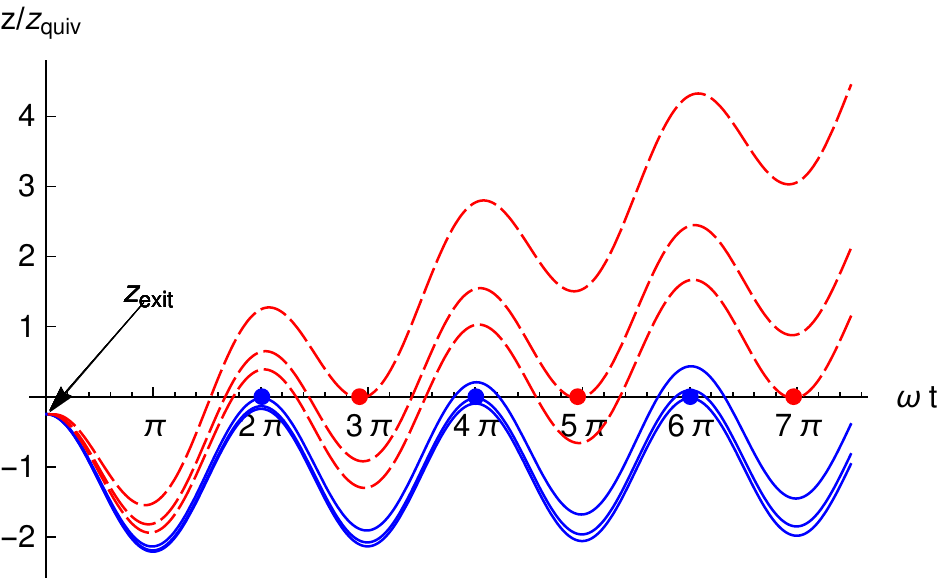}
  \caption{Trajectories with soft recollisions after tunnel ionization, for a Keldysh parameter of  $\gamma=\figureonegamma$.}
  \label{fig:trajectories_at_transitions} 
\end{figure}

This raises an intriguing possibility, because there is in general a discrete sequence of soft-recolliding trajectories \cite{Rost_JPhysB,Becker_rescattering}, shown in \reffig{fig:trajectories_at_transitions}, which spend increasing amounts of time in the continuum before recolliding, and which should appear as distinct peaks in photoelectron spectra. In this work we extend this sequence and show that there are, in fact, two families of soft-recolliding trajectories, which revisit the ion at even or odd numbers of half-periods after ionization. 

The previously-described trajectories \cite{Rost_JPhysB,Becker_rescattering,lemell_lowenergy_2012,lemell_classicalquantum_2013} recollide after an odd number of half-periods, and their drift energy is a constant multiple of the ponderomotive potential $U_p$. In contrast, trajectories with soft recollisions at integral multiples of the laser period have lower energies, of around 1~meV for common experimental parameters, which scale rather unfavourably as $1/U_p$. This series seems to have been overlooked, but the low energies mean that the peaks can (and, in fact, will) contribute to the `zero-energy structure' (ZES) found re\-cent\-ly~\cite{ZES_paper,pullen_kinematically_2014,dura_ionization_2013}.

Moreover, we show that both families of soft recollisions come up naturally within the analytical $R$-matrix theory of photoionization \cite{ARM_initial,ARM_circular, ARM_abinitio_verification, ARM_trajectories,ARM_attoclock,MResReport}.
The ARM theory implements the intuition that the dominant effect of the ion's electrostatic potential on the photoelectron is a phase, and makes this intuition rigorous via the use of eikonal-Volkov wavefunctions \cite{eikonalVolkov_initial,eikonalVolkov_prelim} for the photoelectron. As in SFA treatments, this results in a time integral over the ionization time which, when approximated using saddle-point methods, produces a quantum-orbit picture based on semiclassical trajectories. 

This first-principles approach has the advantage that it provides initial conditions for the trajectories by matching the eikonal-Volkov solution away from the core to the bound wavefunction near the core, using the WKB approximation for the bound wavefunction. In particular, the trajectories produced by this matching procedure are real-valued at the entrance to the classically-forbidden region, and this results in a nonzero imaginary position at real times, after exiting the classically-forbidden region.

Thus, ARM operates with quantum orbits \cite{salieres_quantum_orbits} which, in contrast to their purely real-valued counterparts in the Coulomb-corrected SFA~\cite{CCSFA_initial_short, CCSFA_initial_full,TCSFA_sub_barrier,yan_TCSFA_caustics}, can change the amplitude of their contribution after ionization, i.e., after the tunnelling electron leaves the classically forbidden region \cite{ARM_trajectories,ARM_attoclock}. Mathematically, this change in the amplitude comes through the imaginary part of the position, which results in an imaginary part of the action that directly affects the amplitude. Physically, this mirrors the dynamical focusing effect found in approaches that use the full classical trajectory \cite{Rost_JPhysB,Rost_PRL}.

The importance and the effect of the imaginary part of the semiclassical trajectory is most strongly felt when the electron revisits the core. Mathematically, one needs to extend the core potential into the complex plane. When the real part of the trajectory is small, this analytical continuation has a branch cut which cannot be integrated across. This branch cut must be managed carefully, as it can preclude several standard choices of integration contour, including in particular the contour going from the ionization time directly to the real time axis and then along it. In these cases, one needs a more flexible approach towards contour choice; to ease this choice we briefly describe user-friendly software \cite{QuantumOrbitsDynamicDashboard,QODD_Software_paper} to visualize the effects on the complex-valued position of different contour choices.

Moreover, we present a method for consistently and programmatically navigating the Coulomb branch cuts to calculate photoelectron spectra. This method is based on the fact that the branch cuts come in pairs which always contain between them a saddle point of the semiclassical distance to the origin, $\sqrt{\vbr(t)^2}$; this saddle point is a solution to the closest-approach equation $\vbr(t)\cdot\vbv(t)=0$ on the complex plane. These times of closest approach offer a rich geometry of their own, both with complex quantum orbits and within the more restricted simple man's model. More practically, by choosing appropriate sequences of closest-approach times, one can systematically choose correct integration contours.

Within this framework, soft recollisions appear as complex interactions between different sets of branch cuts, marked by close approaches between two or three closest-approach saddle points and by topological changes in the branch cut configuration. This happens only at very low transverse momenta, and is managed well by our method. Nevertheless, the increased time spent by the electron near the ion -- mirrored in the ARM formalism by close approaches between singularities and saddle points -- results in an increase in the amplitude that reflects the photoelectron peaks seen in experiments.

In the current approach, the Coulomb field of the ion is allowed to influence the phase of the wavefunction (and from there the ionization amplitude), but not the underlying trajectory. A first-principles approach based on the full trajectory is still lacking, but such a theory will likely require the trajectory to be real \textit{before} it enters the classically-forbidden region during the ionization step. It is then likely that such a theory will contain many of the elements we describe for the laser-driven trajectories, including the branch cut behaviour and its navigation. In that sense, our work is a roadmap for those difficulties and their resolution.

This paper is structured as follows. In Section~\ref{sec:classical-transitions} we explore soft recollisions for the classical trajectories of the simple man's model, giving simple approximations for the momenta that produce them, and we explore their scaling. In Section~\ref{sec:ARM-theory} we derive a simple version of ARM theory, emphasizing the features that give rise to complex-valued positions for the quantum orbits.

We then examine, in Section~\ref{sec:temporal-branch-cuts}, the ways in which complex-valued positions give rise to branch cuts and the cases in which the latter make more sophisticated contour choices necessary. In Section~\ref{sec:times-of-closest-approach} we study the geometry of the times of closest approach, both within the simple-man's model and for the ARM quantum-orbit picture, and explain their use for navigating integration contours around the Coulomb branch cuts. We then present the resulting photoelectron spectra in Section~\ref{sec:results} and summarize our results in Section~\ref{sec:conclusions}. In addition, we provide supplementary information \cite{SupplementaryInformation} with interactive 3D versions of several figures in this paper.

\section{Soft recollisions in classical trajectories}
\label{sec:classical-transitions}

The simplest approach to the dynamics of a photoelectron after tunnel ionization is known as the simple man's model. In its quantum version it accounts for the tunnelling process using the strong-field approximation, and then allows the electron to follow a laser-driven trajectory after it exits the tunnelling barrier. Within the SFA, the electron has velocity
\begin{equation}
\vbv(t)=\vbp+\vba(t)
\end{equation}
where $\vbp$ is the canonical momentum registered at the detector. The electron is `born' into the continuum at the origin at a complex starting time $\ts=\tn+i\tauT$ which obeys
\begin{equation}
\frac12\left(\vbp +\vba(\ts)\right)^2+I_p=0,
\end{equation}
where $\vba(t)=-\int\vbf(t)\d t$, $\vbf(t)=F\cos(\omega t)\uz$ is the laser's electric field, which we take to be a monochromatic, linearly polarized pulse, and $I_p=\tfrac12\kappa^2$ is the ionization potential. (We use atomic units unless otherwise noted.) The resulting trajectory is then
\begin{equation}
\rcl(t)=\int_\ts^t\left(\vbp+\vba(\tau)\right)\d\tau,
\end{equation}
which is in general complex-valued.

Throughout this work, we understand a `soft recollision' to mean a trajectory which has a laser-driven turning point, with essentially zero velocity, in the neighbourhood of the ion \cite{Rost_PRL,Rost_JPhysB}. This can occur for a range of impact parameters, and the most relevant trajectories will pass within a few bohr of the ion. However, it is often easier to identify those \textit{laser-driven} trajectories that `go through' the ionic core, so we focus on them in the understanding that the neighbouring trajectories are the most relevant. 

Thus, to look for soft recollisions in the corresponding classical trajectories, we require that the real parts of both the velocity and the position vanish:
\begin{subequations}
\label{symbolic_system}
\begin{empheq}[left=\empheqlbrace]{align}
\Re\left[\zcl(\tr)\right]&=\Re\left[ \int_\ts^\tr \left(p_z+A(\tau)\right)\d\tau\right]=0 \\
v_z(\tr)&=p_z+A(\tr)=0.
\end{empheq}
\end{subequations}
This can be expressed as
\begin{subequations}
\label{spelled_out_system}
\begin{empheq}[left=\empheqlbrace]{align}
\zexit+p_z(\tr-\tn)+\frac{F}{\omega^2}\left(\cos(\omega\tr)-\cos(\omega\tn)\right)  &=0 \\
p_z-\frac F\omega \sin(\omega\tr)  &=0,
\end{empheq}
\end{subequations}
where 
\begin{align}
\zexit
&=
\Re\left[ \int_\ts^\tn \left(p+A(\tau)\right)\d\tau\right]
\nonumber\\ &=
\frac{F}{\omega^2}\cos(\omega\tn) \left(1- \cosh(\omega\tauT)\right)
\end{align}
is known as the tunnel exit.

This system can be solved numerically, but it is more instructive to consider its linearized version with respect to $p_z$, since all the soft recollisions happen at small energies with respect to $U_p$. To do this we express the starting time as
\begin{align}
\tn+i\tauT=\ts
& = \frac1\omega  \arcsin\left(\frac{\omega}{F}(p_z+i\kappa)\right)
\nonumber\\& \approx  \frac{ p_z}{F}\frac{1}{\sqrt{1+\gamma^2}} + \frac{i}{\omega}\arcsinh\left(\gamma\right),
\end{align}
where $\gamma=\omega\kappa/F$ is the Keldysh parameter, so that
\begin{equation}
\zexit\approx - \frac{F}{\omega^2}\left(\sqrt{1+\gamma^2}-1\right).
\end{equation}
In the tunnelling limit of $\gamma \ll 1$, $\zexit$ reduces to $-I_p/F$ as expected. 

\begin{figure}
  \includegraphics[width=\columnwidth]{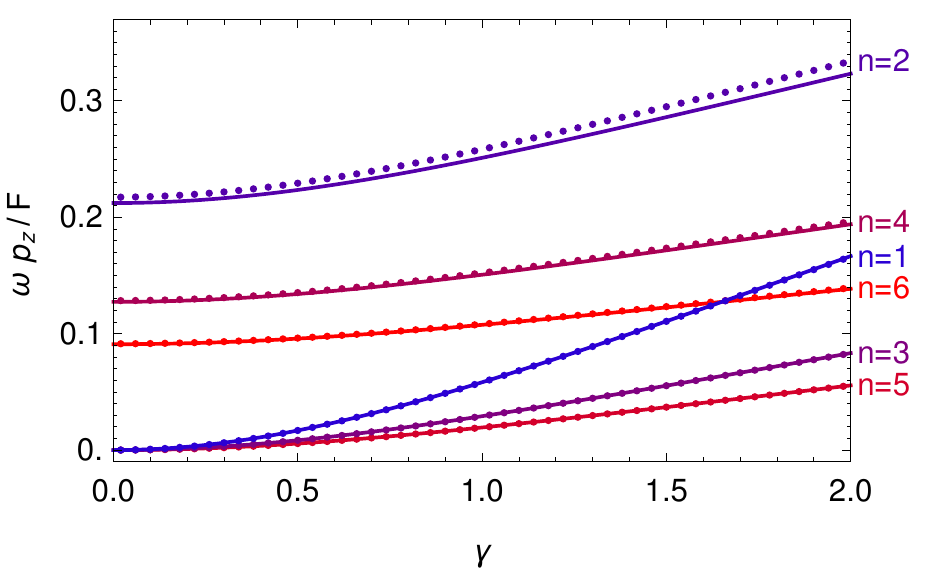}
  \caption{Scaling of the normalized momenta $\omega p_z/F$ as a function of the Keldysh parameter $\gamma$ for the first six soft-recollision trajectories. Dots show the exact solutions of \eqref{symbolic_system} and lines show the linearized result \eqref{linearized_momenta}.}
  \label{fig:scaling}  
\end{figure}

The linearized system now reads
\begin{subequations}
\begin{empheq}[left=\empheqlbrace]{align}
p_z\tr+\frac{F}{\omega^2}\left(\cos(\omega\tr)-\sqrt{1+\gamma^2}\right)  &=0 \\
p_z-\frac F\omega \sin(\omega\tr)  &=0,
\label{pz-to-tr-eqn}
\end{empheq}
\end{subequations}
and to obtain a solution we must linearize $\tr$ with respect to $p_z$. Examination of the numerical solutions of the original system shows that soft recollisions exist for every half period after $\omega\tr=2\pi$, so we write
\begin{equation}
\omega \tr= (n+1)\pi+\omega \,\delta\tr
\end{equation}
with $n=1,2,3\ldots$ indexing the solutions. With this we obtain from \eqref{pz-to-tr-eqn} that $\delta\tr\approx(-1)^{n+1}p_z/F$ and $\cos(\omega\tr)\approx(-1)^{n+1}$. This gives in turn the drift momentum of the successive soft-recolliding trajectories as 
\begin{equation}
\pzsr \approx \frac F\omega \frac{\sqrt{1+\gamma^2}+(-1)^n}{(n+1)\pi}.
\label{linearized_momenta}
\end{equation}
These are shown in \reffig{fig:scaling}, and are generally a good approximation to the exact solution of the system \eqref{symbolic_system}.

It is clear here that trajectories with odd and even $n$, which approach the origin from different sides, will behave very differently, and in particular will have very different scaling for low $\gamma$. In particular,
\begin{itemize}
  \item
    for \odd\ trajectories, in the tunnelling limit
    \begin{equation}
      \pzsr 
      \approx \frac F\omega \frac{\gamma^2}{(n+1)2\pi}
      =\frac{\gamma\kappa}{(n+1)2\pi},
    \label{pzsr-linearized-odd-n}
    \end{equation}
    and the kinetic energy scales as $\gamma^2$ for a fixed target species, whereas
  \item
    for \even\ trajectories
    \begin{equation}
      \pzsr
      \approx \frac F\omega \frac{2}{(n+1)\pi},
    \end{equation}
    and the kinetic energy is a constant fraction of $U_p$, as found previously \cite{Rost_PRL,Rost_JPhysB}. This scales as $\gamma^{-2}$.
\end{itemize}

These differences are grounded in much simpler scaling considerations. For \even\ trajectories, the centroid of the oscillation must advance past the ion by twice the quiver radius $\zquiv=F/\omega^2$ plus the tunnel length $\zexit =\tfrac{1}{2}\gamma^2\, \zquiv$ within the given number of laser periods, which implies a scaling of the form
\begin{equation}
\pzsr\sim\frac{2\zquiv+\zexit}{T}\approx\frac{2F/\omega^2}{2\pi/\omega}\sim\frac F\omega,
\end{equation}
and an energy which scales as $\gamma^{-2}$. (Further, this suggests that the small observed deviations from this exponent \cite{blaga_original_LES, LES_Scaling} could also be explained by a shifted power law to account for the tunnel exit, as confirmed in \citer{murnane_TCSFA_tunnel_exit}.) For the \odd\ trajectories, on the other hand, the centroid must only advance by the tunnel length in a multiple of a laser periods, and the scaling reflects this:
\begin{equation}
\pzsr\sim\frac{\zexit}{T}=\frac{I_p/F}{2\pi/\omega}\sim\frac{I_p\omega}{F}.
\end{equation}
Thus, the fixed distance to cover in an increasing amount of time gives these trajectories their very low momenta at long wavelengths. Indeed, for argon in a \SI{3.1}{\micro\meter} field of intensity \SI{e14}{W/cm^2}, as in  \citer{pullen_kinematically_2014}, the highest \odd\ momentum is $p_z\approx \SI{0.02}{\au}$ This corresponds to a minimum energy of \SI{8}{meV}, which is consistent with the observed momentum spread of the ZES.


The difference in scaling behaviour also suggests avenues of experimental research for testing this connection, as well as for probing the internal structure of the ZES. In particular, it's desirable to have experiments which increase the energy scale of the ZES and lift it above the consistent-with-zero level of the current experimental resolution, so that its details can be better examined. If it is caused by soft-recolliding trajectories of this nature, then its energy should scale as $\gamma^2 I_p$, which points the way to experiment.

This scaling is fairly unfavourable, since it is also necessary to keep $\gamma$ small to remain in the tunnelling limit. This means, then, that the ZES is best probed using high-$I_p$ targets, of which the most natural is $\mathrm{He}^+$ ions -- either as an ionic beam or prepared locally via sequential ionization or a separate pre-ionizing pulse. This brings a four-fold increase in $I_p$ compared to hydrogen, and allows for a wider range of intensities and wavelengths which keep $\gamma=\kappa\omega/F$ small but $\pzsr\sim\gamma\kappa=\kappa^2\omega/F$ large enough to be directly measured.

Finally, we note that the momentum ratios between the two families of trajectories form universal sequences independent of the system's parameters: $3/5,5/7,7/9,\ldots$ for \even\ trajectories and $1/2,2/3,3/4,\ldots$ for \odd\ ones. The latter sequence should be difficult to resolve using current experimental setups but, if observed, would allow a richer window with which to observe the tunnel exit distance~\cite{murnane_TCSFA_tunnel_exit}.

It is also important to point out that, irrespective of the precise mechanism which translates the soft-recolliding trajectories into peaks in the photoelectron spectrum, this mechanism should apply equally well to both series of trajectories. As pointed out above, quantum-orbit approaches increase their associated amplitude via imaginary parts of the action, whereas classical-trajectory approaches give photoelectron peaks through dynamical focusing near the soft recollisions. In either case, the dynamical similarity between the two series of recollisions, evident in \reffig{fig:trajectories_at_transitions}, should cause similar results for both. The reason the \odd\ trajectories have not appeared explicitly so far seems to be only their very low energy.%
\footnote{
In addition, if the position of the tunnel exit is dropped from the analysis then all \odd\ trajectories will degenerate to zero energy, which helps confound the situation.
}

\section{Analytical \texorpdfstring{$R$}{R}-matrix theory}
\label{sec:ARM-theory}
We will now give a brief recount of ARM theory, as developed in \citers{ARM_initial,ARM_circular}, emphasizing the features that provide initial conditions for the complex-valued quantum orbit trajectories, thereby leading to branch cuts in the complex time plane.

The $R$-matrix approach was adapted from nuclear phy\-sics~\cite{Rmatrix_nuclear} to model collisions of electrons with atoms~\cite{Rmatrix_atomic} and molecules~\cite{Rmatrix_molecular}, and it is designed to deal with electron correlation effects by confining full many-body dynamics to the interior of a sphere around the system, where they are most important, and using a single-active-electron approximation outside it. In a strong-field context, this permits a fuller account of multi-electron effects \cite{Rmatrix_numerical,ARM_initial_multielectron}, and it also allows a smoother matching between the WKB asymptotics of the bound states near the ion with the perturbing effect of the ion's potential on the continuum wavefunction on the outside region where this effect is small.

The effect of the ion can be accounted for rigorously, in the outer $R$-matrix region, by using the eikonal approximation \cite{eikonalVolkov_prelim, eikonalVolkov_initial} on the usual Volkov states, where the phase and amplitude of the wavefunction are approximated by the first terms in a semiclassical series in powers of $\hbar$. This approximation yields continuum time-dependent wavefunctions of the form
\begin{align}
\bracket{\vbr}{\peva(t)}
= &
\frac{1}{(2\pi)^{3/2}}
e^{
  i(\vbp+\vba(t))\cdot\vbr
  -\frac i2 \int_T^t\left(\vbp+\vba(\tau)\right)^2\d\tau
}
\nonumber\\ & \quad\qquad\qquad \times
e^{ -i\int_T^t U(\rl(\tau;\vbr,\vbp,t))\d\tau },
\label{EVA-wavefunctions}
\end{align}
where $U$ is the ion's electrostatic potential, and 
\begin{equation}
\rl(\tau;\vbr,\vbp,t)=\vbr + \int_t^\tau \left(\vbp+\vba(\tau')\right) \d\tau'
\label{laser-driven-trajectory}
\end{equation}
is the laser-driven trajectory that starts at position $\vbr$ at time $t$ and has canonical momentum $\vbp$. These eikonal-Volkov states are (approximate) solutions of the time-dependent Schr\"odinger equation with the hamiltonian 
\begin{equation}
H=\tfrac12 \vbp^2 + U(\vbr) +\vbr\cdot\vbf,
\end{equation}
or, in other words, they are propagated in time as
\begin{equation}
\ket{\peva(t)}=U(t,t')\ket{\peva(t')}\text{ for }t>t'
\end{equation}
where the propagator $U(t,t')$ obeys the (approximate) Schr\"odinger equation
\begin{equation}
i\frac{\partial}{\partial t}U(t,t')= H U(t,t')\text{, with } U(t,t)=1.
\end{equation}

The inner and outer regions are separated by a sphere around the ion of radius $a$. In strong fields it will be required to satisfy the conditions $1/\kappa \ll a \ll I_p / F$, so it is well away from both ends of the tunnelling barrier; the wavefunction is correspondingly split into an inner and outer wavefunctions. This carries the problem that the hamiltonian is no longer hermitian in either region (as the usual integration-by-parts proof leaves uncancelled boundary terms), so the resulting system has two non-hermitian Schr\"odinger equations coupled through their boundary conditions. 

The problem is rigorously addressed by introducing a `hermitian completion' of the hamiltonian, by means of the Bloch operator
\begin{equation}
L=\delta(r-a)\left(\frac{\partial}{\partial r} +\frac{1-b}{r}\right),
\end{equation}
for which $H+L$ is hermitian in the inner region and $H-L$ is hermitian in the outer region, with the Bloch operator cancelling out the boundary terms in the usual integration by parts. The system can then be re-expressed as two coupled, hermitian, inhomogeneous Schr\"odinger equations:
\begin{subequations}
\begin{align}
i\frac{\partial}{\partial t}\Psi_\mathrm{in}(\vbr,t)
& = 
 (H+L)\Psi_\mathrm{in}(\vbr,t)-L\Psi_\mathrm{out}(\vbr,t)
\\
i\frac{\partial}{\partial t}\Psi_\mathrm{out}(\vbr,t)
& = 
 (H-L)\Psi_\mathrm{out}(\vbr,t)+L\Psi_\mathrm{in}(\vbr,t).
 \label{Bloch-Schrodinger-equation}
\end{align}
\end{subequations}
In particular, since the Bloch term $L\Psi$ is local to the boundary, we can implement the boundary conditions by substituting the wavefunction from the other side of the boundary in both equations, which then acts as a source term. The solutions to the coupled equations then automatically obey the boundary conditions.

Once with a well-formulated system, one can apply the relevant approximations. In practice, the equation for the inner region is not affected as long as the ionization rate is not too great, since then the Bloch term $L\Psi_\mathrm{out}$ can be neglected, or its main effects can be modelled by incorporating a ground-state depletion factor; furthermore, by an appropriate choice of the constant $b$ in the Bloch operator, the ground state $\Psi_g$ can be made an eigenstate of $L$. (In particular, choosing $b=1/\kappa$ implies that $L\Psi_g=-\kappa\delta(r-a)\Psi_g$.) Similarly, the Bloch term in the outer region's hamiltonian can be neglected as the ionized wavefunction will generally be far from the boundary.

Given these approximations, one can then write down the formal solution to the outer Schr\"odinger equation \eqref{Bloch-Schrodinger-equation} as
\begin{equation}
\ket{\Psi(t)}= -i \int_{-\infty}^t U(t,t') L \ket{\Psi_\mathrm{in}(t')}\d t'.
\end{equation}
The quantity of interest is the photoelectron momentum amplitude at a time $T$ long after the pulse is finished, which is then
\begin{align}
a(\vbp)
& = 
\bracket{\vbp}{\Psi(T)}
= -i \int_{-\infty}^T \bra{\vbp}U(T,t) L \ket{\Psi_\mathrm{in}(t)}\d t
\nonumber \\ & =
-i \int_{-\infty}^T \bra{\peva(t)}L \ket{\Psi_g(t)}\d t
.
\end{align}
To calculate the spatial matrix element, we work in the position basis, so
\begin{align}
a(\vbp)
& = 
-i \int\d\vbr \int_{-\infty}^T\!\!\!\!\d t
\bracket{\peva(t)}{\vbr} \matrixel{\vbr}{L}{\Psi_g(t)}
\nonumber \\ & =
-i \int\d\vbr \int_{-\infty}^T\!\!\!\!\d t\,
(-\kappa)\delta(r-a)\bracket{\vbp+\vba(t)}{\vbr} \bracket{\vbr}{\Psi_g}
\nonumber \\ & \qquad \times
e^{iI_pt +\frac i2 \int_T^t\left(\vbp+\vba(\tau)\right)^2\d\tau}
e^{ i\int_T^t U(\rl(\tau;\vbr,\vbp,t))\d\tau }
\nonumber \\ & =
-i\int_{-\infty}^T\!\!\!\!\d t\,
e^{iI_pt +\frac i2 \int_T^t\left(\vbp+\vba(\tau)\right)^2\d\tau}
\nonumber \\ & \qquad \times
(-\kappa)\int\d\vbr 
\delta(r-a)\bracket{\vbp+\vba(t)}{\vbr} \bracket{\vbr}{\Psi_g}
\nonumber \\ & \qquad \times
e^{ i\int_T^t U(\rl(\tau;\vbr,\vbp,t))\d\tau }
.
\label{initial-temporal-integral}
\end{align}

This calculation produces, then, an ionization amplitude $a(\vbp)$ in the form of a temporal integration of a phase factor
\begin{equation}
e^{iS_V(t)}=\exp\left[i\left(I_pt +\frac 12 \int_T^t\left(\vbp+\vba(\tau)\right)^2\d\tau\right)\right],
\label{SFA-phase-factor}
\end{equation}
a Coulomb phase factor arising from the integration of the ionic potential along the laser-driven trajectory, and a spatial factor involving a Fourier transform of the ground state. This amplitude is structurally similar to the SFA amplitude: in particular, the phase factor \eqref{SFA-phase-factor} is present in the SFA. It oscillates at the frequencies $I_p$ and $U_p$, which are fast compared to the laser-cycle timescales that the integration takes place on, and this makes the integral highly oscillatory. This enables, in turn, the application of the saddle-point approximation \cite{Bleistein_Integrals}. However, this is complicated by the fact that the eikonal Coulomb correction to the Volkov phase,
\begin{equation}
\exp\left[ i\int_T^t U(\rl(\tau;\vbr,\vbp,t))\d\tau \right]
\label{Coulomb-correction-phase}
\end{equation}
couples the spatial and temporal integrations in a complicated fashion. This factor, moreover, is crucial in ensuring that the resulting amplitude is independent of the (unphysical) boundary radius $a$.

To disentangle both integration steps, one first performs the temporal saddle-point integration, with an $\vbr$-dependent result since the phase now includes a spatial term $e^{i\vba(t)\cdot\vbr}$. The resultant saddle point $t_a$ is generally not far from the standard SFA saddle $\ts=\tn+i\tauT$, which obeys
\begin{equation}
\frac 12 \left(\vbp+\vba(\ts)\right)^2+I_p=0.
\label{SFA-saddle-point-equation}
\end{equation}
Moreover, the variations in $t=t_a$ in the Coulomb correction term \eqref{Coulomb-correction-phase} exactly cancel out the variation with respect to $a$ in the ground-state wavefunction \cite[see][]{ARM_initial,ARM_circular,MResReport}.

The result of this matching procedure \cite{ARM_initial,ARM_circular,MResReport} is a simple expression for the ionization amplitude with most factors evaluated at the complex saddle-point time~$\ts$ of \eqref{SFA-saddle-point-equation},
\begin{align}
	a(\vb{p})
	& = 
    e^{i I_p \ts}
	e^{-\frac{i}{2} \int_{\ts}^T\left(\vbp+\vba(\tau)\right)^2\d\tau}
\nonumber \\ & \qquad \times
	e^{-i\int_{t_\kappa}^T U\left(\int_{ \ts}^\tau \vbp+\vba(\tau')\d\tau'\right) \d\tau}
	R(\vbp)
	,
\label{ARM-final-result}
\end{align}
where $\tk=\ts-i/\kappa^2$ and the final amplitude will be a sum over all saddle points $\ts$ with positive imaginary part. (Hereafter, however, we focus on the contribution from a single half-cycle.) In this expression the spatial factors have been incorporated into a shape factor
\begin{align}
	R(\vbp)  =
	\frac{i \kappa a^2 }{\sqrt{iS_V''(\ts)}} 
	&
	\int\frac{\d\Omega}{2\pi}
	e^{-i\int_{t_a}^{t_\kappa}U\left(\int_{\ts}^\tau \vbp+\vba(\tau')\d\tau'\right) \d\tau}
\nonumber \\ & \qquad \times
	e^{-i\left(\vbp+\vb{A}(\ts)\right)\cdot\vb{r}}
	\bracket{\vbr}{\Psi_g}|_{r=a} 
	,
\end{align}
which is independent of $a$ \cite{ARM_initial,ARM_circular,MResReport}. This shape factor is essentially the Fourier transform of the ground state over the spherical boundary; in this sense it is analogous to the planar boundary used in partial Fourier transform approaches \cite{Partial_Fourier_transform}.

\begin{figure}[b]
  \includegraphics[width=0.8\columnwidth]{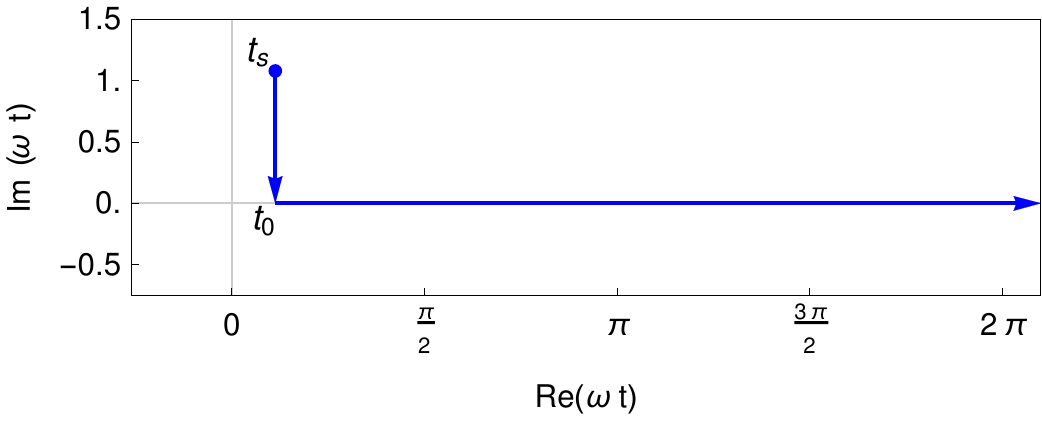}
  \caption{The time contour for integrating the exponents in Eq.~\eqref{ARM-final-result} is usually taken from $\ts$ down to its real part $\tn$ and then along the real axis until the detection time $T$.}
  \label{fig:standard-time-contour}
\end{figure}

The ARM ionization amplitude Eq.~\eqref{ARM-final-result} is similar to the SFA result, and it admits a clear quantum-orbit trajectory interpretation \cite{salieres_quantum_orbits}. The electron is ionized at time $\ts$ and propagates to its detection, at real and large time $T$, with velocity $\vbp+\vba(t)$ and position 
\begin{equation}
\rcl(t)=\int_{\ts}^t \left(\vbp+\vba(\tau)\right) \d\tau.
\label{complex-trajectory}
\end{equation}
Along the way, it acquires `phases' given by the exponentiated action of the different relevant energies: $e^{i I_p \ts}$ from the bound energy before ionization, $e^{-\frac{i}{2} \int_{\ts}^T\left(\vbp+\vba(\tau)\right)^2\d\tau}$ from the kinetic energy after ionization, and $e^{-i\int_{t_\kappa}^T U\left(\rcl(\tau)\right) \d\tau}$ from the Coulomb interaction. It is important to remark that these factors are no longer pure phases since the exponents are now complex, which greatly affects their amplitude and indeed completely encodes the (un)likelihood of the tunnelling process.

The complex integrals in Eq.~\eqref{ARM-final-result} are complex contour integrals which can be taken, in principle, over any contour which joins both endpoints. In practice, the standard contour goes directly down from $\ts$ to its real part $\tn$ and then along the real axis until the detection time $T$, as shown in \reffig{fig:standard-time-contour}. This conveniently separates the imaginary part of the contour, which yields an amplitude reduction $\left|e^{i I_p \ts} e^{-\frac{i}{2} \int_{\ts}^{\tn}\left(\vbp+\vba(\tau)\right)^2\d\tau}\right|$ that encodes the tunnelling probability, and the normal propagation along the real time axis. As we shall see later, when the Coulomb interaction is included over one or more laser periods, this choice  of the second leg of the contour is no longer necessarily ideal or even allowed.

The main effect of the matching procedure is that the $\vbr$-dependent laser-driven trajectory \eqref{laser-driven-trajectory} has been replaced by a classical trajectory of Eq.~\eqref{complex-trajectory} starts at the origin, but this trajectory is only integrated over, in the Coulomb correction factor 
\begin{equation}
\exp\left[-i\int_{t_\kappa}^T U\left(\rcl(\tau)\right) \d\tau\right],
\label{final-correction-factor}
\end{equation}
from a time $\tk=\ts-i/\kappa^2$ just under the saddle point. This shift arises here rigorously as a result of matching the outer-region wavefunction to the WKB asymptote of the bound wavefunction. Since its origin is intrinsically spatial, it is preferable to other regularization procedures such as the ionization-rate-based scheme of \citer{LES_Scaling}.

Finally, it is important to note that the process that leads to the Coulomb correction factor \eqref{final-correction-factor} is little more than the analytic continuation of the original factor \eqref{Coulomb-correction-phase}, which appeared in the initial real-valued temporal integral \eqref{initial-temporal-integral}. This uniquely determines the trajectory to use for the Coulomb correction, with the initial conditions set by the matching procedure. The trajectory $\rcl(t)$, along which the Coulomb correction is integrated, obeys the laser-only equation of motion,
\begin{equation}
\ddot{\vbr}_{\cl}(t)=-\vbf(t), 
\end{equation}
as opposed to the full equation (including both the laser and the Coulomb field) used in the CCSFA approach \cite{CCSFA_initial_short, CCSFA_initial_full,TCSFA_sub_barrier,yan_TCSFA_caustics}. This is a direct consequence of the appearance of $\rcl$ in the EVA wavefunction~\eqref{EVA-wavefunctions}. Work to include the effect of the ion on the trajectory should ideally be directed at its inclusion at the level of this wavefunction.

The initial conditions are also different to those used in the CCSFA approach. In ARM theory, the initial position is obtained from first principles and it is real at the entrance of the tunnelling barrier instead of its exit, as is imposed externally on the CCSFA trajectory. The consequences of this can be dramatic, as we show below, and among other things they preclude certain choices of integration contour in \eqref{final-correction-factor}. In this light, calculating semiclassical trajectories for the full equation of motion with the correct initial conditions in complex time and space appears to be a far more difficult task.

In the next section we describe the origin of these difficulties, the mathematical challenges that arise, and how they can be addressed within the ARM theory.

\section{Emergence of temporal branch cuts}
\label{sec:temporal-branch-cuts}
The analytical $R$-matrix theory allows us, then, a tra\-jec\-to\-ry-based approach for accounting for the effect of the ionic potential on the photoelectron. Here the potential is integrated over the electron's complex-valued trajectory \eqref{complex-trajectory} to give the correction factor \eqref{Coulomb-correction-phase}, which now has a nontrivial amplitude. The trajectory starts at a real position, $\rcl(\tk)\approx -\uz/\kappa$, but from then onwards both the integrand and the integration time are complex, so the position is therefore complex.

\begin{figure}[ht]
  \includegraphics[width=0.85\columnwidth]{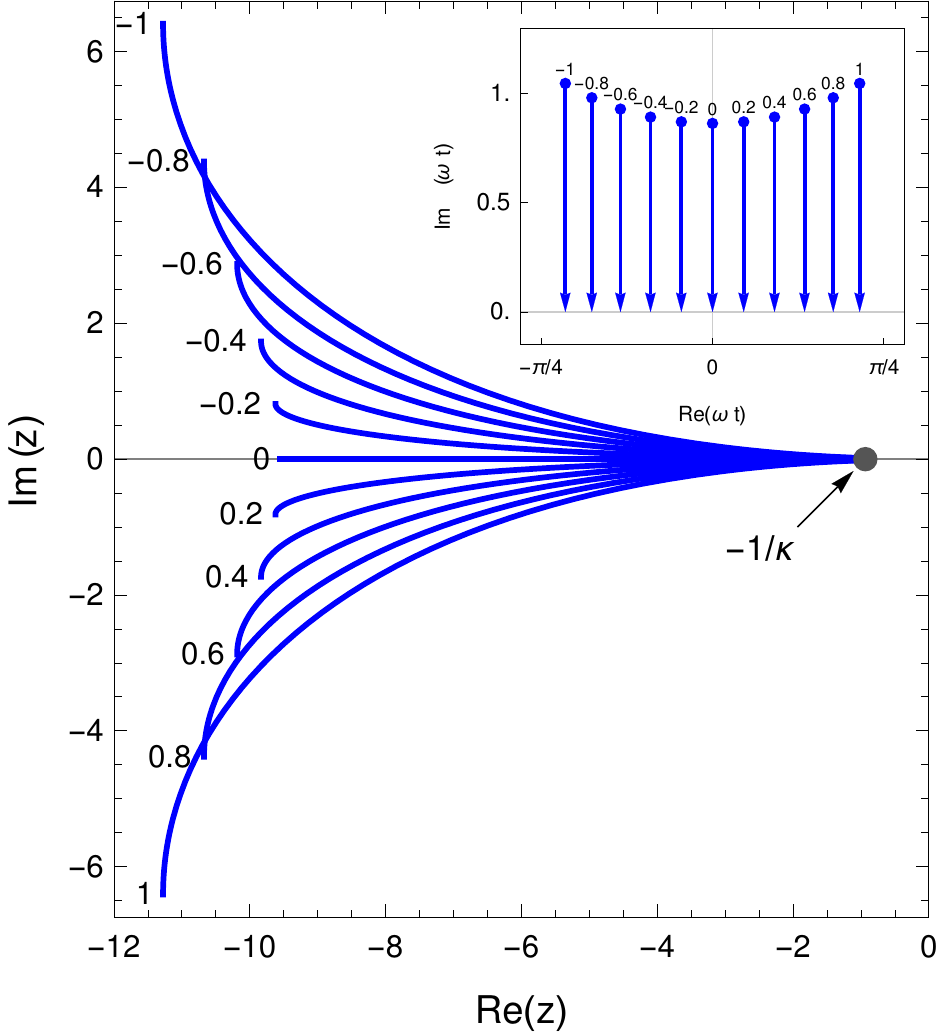}
  \caption{Semiclassical trajectories on the complex $z$ plane, where $t$ covers the `under-the-barrier' segment from $\tk$ to $\tn$, shown in the inset. When $p_z=0$ the trajectory is completely real, but for nonzero momenta the imaginary part is increasingly important.
  Paths are labelled by the normalized longitudinal momentum $\omega p_z/F$. 
  Hereafter $\kappa=\onemicronparskappa$ and $F=\onemicronparsfield$, corresponding to argon in a $\SI{9E13}{\watt/\cm^2}$ field, and $\lambda=\SI{\onemicronparswavelength}{\micro\meter}$ (so $\gamma=\onemicronparsgamma$) unless otherwise noted. }
  \label{fig:complex-z-trajectories}
\end{figure}

The imaginary part of the position has two main sources. The first is along the transverse direction, since any amount of real-valued transverse momentum $\vbp_\perp$ will accumulate an imaginary transverse position $i\vbp_\perp\tauT$ between the complex-valued $\ts=\tn+i\tauT$ and any real time $t$, from the `tunnel exit' $t=\tn$ onwards. The longitudinal position, on the other hand, also obtains imaginary components through the integral of the vector potential, which is now explicitly complex valued, so the end result is the nonlinear dependence shown in \reffig{fig:complex-z-trajectories}.

\begin{figure*}
\centering
\begin{tabular}{ccl}
\raisebox{47mm}{\subfigure[]{\makebox[5mm][0mm]{} }}\hspace{-9mm} &
\addtocounter{subfigure}{-1}
\subfigure{
  \includegraphics[width=1.7\columnwidth]{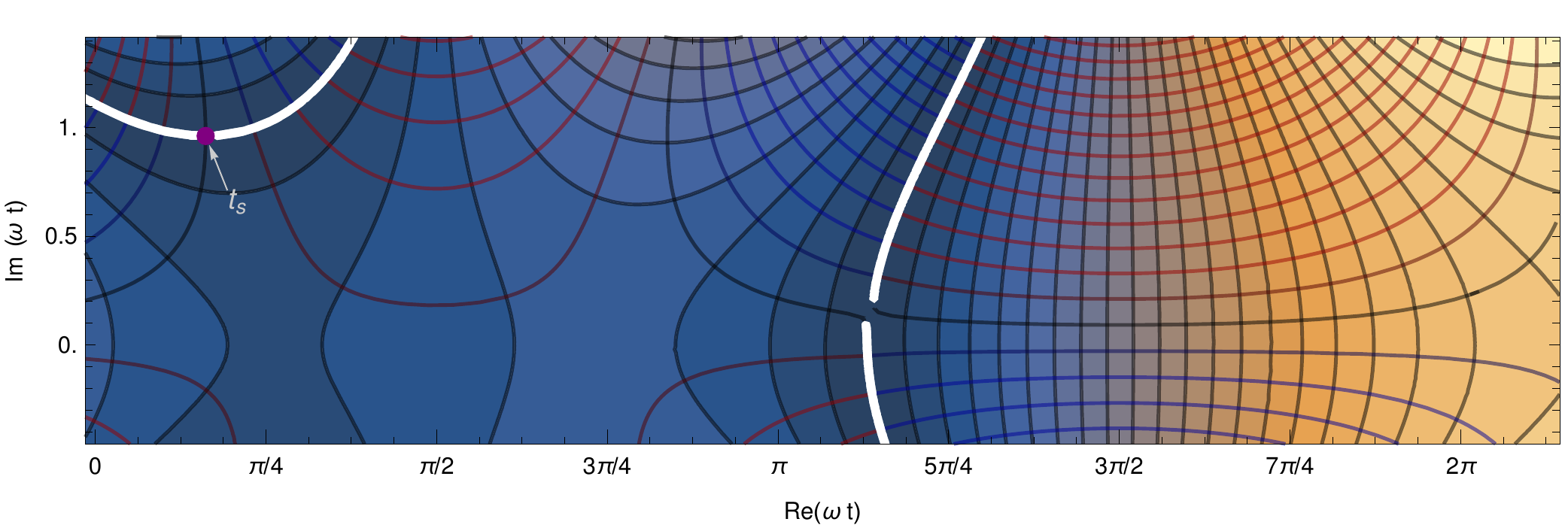}
  \label{fig:complex-position-contour-plot}
  }
\hspace{9mm}
&
  \hspace{-15mm}
  \rotatebox{90}{
    \begin{tabular}{ccc}$\,\,\,\,\,\,\,\,$&
      \rotatebox{-90}{\includegraphics[scale=1]{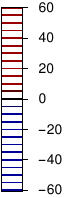}}
      &
      \hspace{-2mm}
      \rotatebox{-90}{\includegraphics[scale=1]{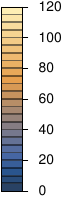}}
      \vspace{1mm}\\\hspace{21pt}& $\scriptscriptstyle \Im \left(\sqrt{\rcl(t)^2}\right)$ & $\scriptscriptstyle \Re\left(\sqrt{\rcl(t)^2}\right)$
    \end{tabular}
  }
\\
\raisebox{47mm}{\subfigure[]{\makebox[5mm][0mm]{}}}\hspace{-9mm} 
\addtocounter{subfigure}{-1}
&
\subfigure{
  \includegraphics[width=1.7\columnwidth]{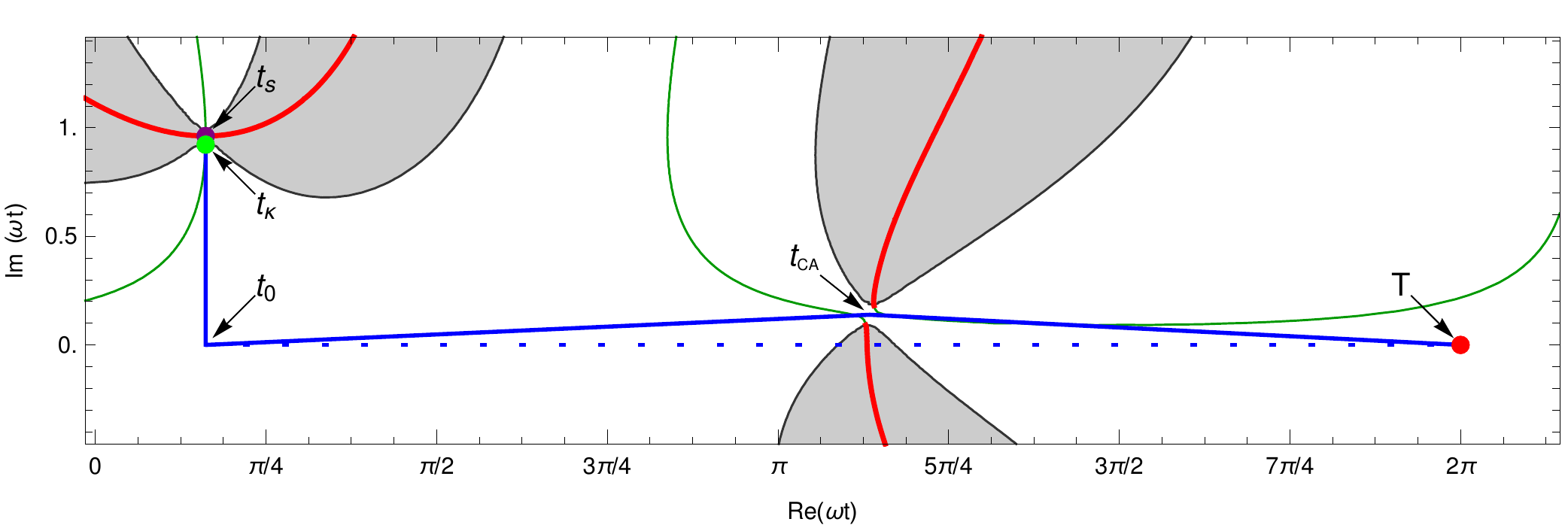}
  \label{fig:branch-cut-sketch}
  }
\hspace{9mm}
& 
\end{tabular}
  \caption{
  \subref{fig:complex-position-contour-plot} Contour plot of the complex distance to the origin, $\sqrt{\rcl(t)^2}$, with the coloured background along the real part and the red, black and blue orthogonal lines representing, respectively, positive, zero and negative imaginary parts. The white lines are branch cuts where the real part is zero and the imaginary part discontinuously changes sign.
  In \subref{fig:branch-cut-sketch} we show a simplification of this picture, with the red lines showing the branch cuts, and the thin green lines the positive real axis of $\vbr^2$. The shaded regions indicate the complex times for which $\Re(\rcl(t)^2)$ is negative, which are undesirable when using gaussian and numerically-obtained ionic potentials, as they would be unphysical there.
  The momentum displayed, $\vbp=(\figurefivepo,0,\figurefivepp)$, is such that the standard contour (integrating down to the real part $\tn$ of the saddle point $\ts$ and then along the real axis, shown dotted in \subref{fig:branch-cut-sketch}) crosses a branch cut. Instead, one should choose a contour which passes through a time between the branch cuts, which we label $\tca$ and explore in detail in Sec.~\ref{sec:times-of-closest-approach}.
}
  \label{fig:complex-position-branch-cuts}
\end{figure*}

\begin{figure}[t]
  \includegraphics[width=\columnwidth]{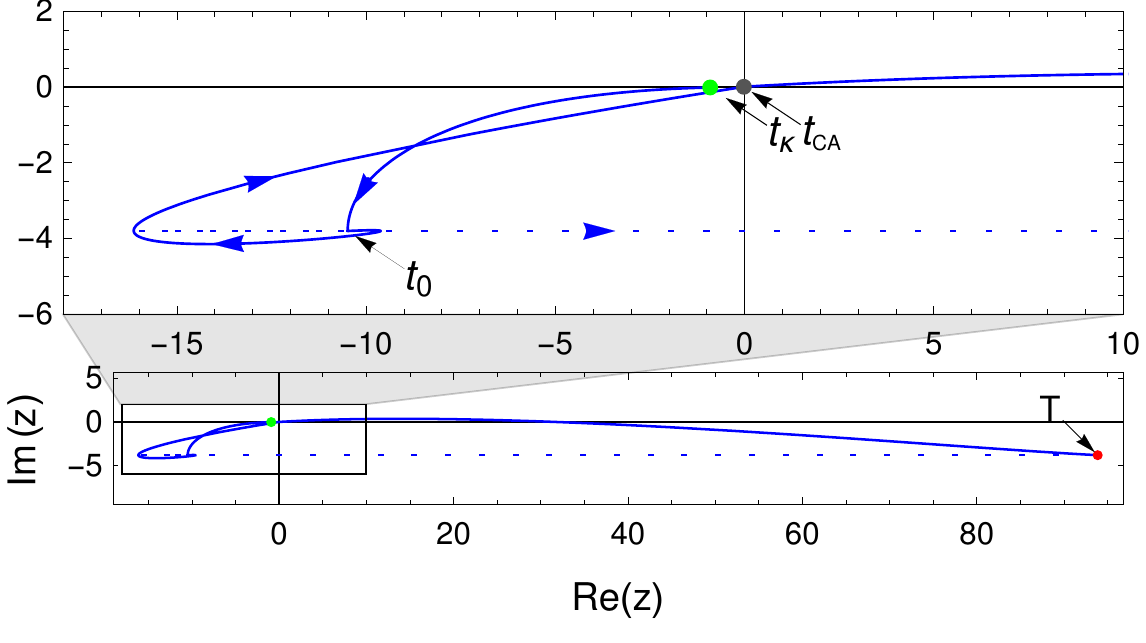}
  \caption{
  Trajectories in the complex $z$ plane corresponding to the contours shown in \reffig{fig:branch-cut-sketch}.   The trajectory starts at $z=-1/\kappa$ at time $\tk$, and departs somewhat from the real position axis as the time goes down to the real axis at $\tn$, as shown in \reffig{fig:complex-z-trajectories}. Along the real axis, on the standard contour shown dashed, the electron goes to large negative $\Re(z)$ before turning around towards the core. It then reaches the ion with a large imaginary part, which causes a discontinuous jump in the square root of \eqref{Coulomb-potential}. Deforming the contours to avoid the branch cuts, shown in the solid line, minimizes the imaginary part of $z$ at the moment of recollision; it is then slowly regained before detection at a large real time $T$. The closest approach time $\tca$ marks the minimum value of $\Re(\rcl(t)^2)$ once the $x$ coordinate is taken into account; there $\zcl$ is small but nonzero.
  }
  \label{fig:complex-z-curved-contours}
\end{figure}

Since the integrand in \eqref{complex-trajectory} is real for real times, this imaginary position will not change along the real time axis, so it will remain present until the electron is detected at a large time $T$, at which the real position will be much larger. During a recollision, however, the real position becomes small. At this time, the imaginary coordinate dominates the position, which becomes an issue when calculating the ion's electrostatic potential. The simplest choice for that is the Coulomb potential,
\begin{equation}
U(\vbr) = -\frac{1}{\sqrt{\vbr^2}}= -\frac{1}{\sqrt{x^2+y^2+z^2}},
\label{Coulomb-potential}
\end{equation}
which is analytical in this form. Other potentials are also tractable, as long as they are analytical functions on the real axis, but pose similar mathematical challenges.\footnote{
In particular, any atomic or molecular charge density is acceptable as long as an explicit analytical formula is available, which in practice is excessively restrictive. Extended spherically-symmetric charge densities like exponentials or softened Coulomb potentials exhibit similar behaviour but with the branch cut pushed back by the characteristic width of the distribution. In contrast, the potential created by a gaussian charge shows drastically different behaviour, with the branch cuts replaced by growth as $e^{+|\vbr|^2}$ where $\Im(\vbr^2)<0$. If the ionic charge distribution is only known numerically -- e.g. via quantum chemical methods -- then both models pose conflicting demands, and numerical evaluation via integration of a Coulomb kernel fails to give analytical results. We will analyse these differences in detail in a future publication.
} For the Coulomb potential, the extension procedure requires the use of a square root, which in general poses at most an integrable singularity, but for complex arguments it has a branch cut along the ray $\vbr^2\in(-\infty,0)$ which cannot be ignored.

This branch cut is reached, precisely, on recollisions where the imaginary part dominates a nonzero real part. Indeed, the squared position can be decomposed as
\begin{equation}
\vbr^2=\Re(\vbr)^2-\Im(\vbr)^2+2i\Re(\vbr)\cdot\Im(\vbr),
\end{equation}
which means that, when $\Im(\vbr)^2>\Re(\vbr)^2$, the trajectory will reach the branch cut at the point where the projection $\Re(\vbr)\cdot\Im(\vbr)$ changes sign. This is marked by a discontinuous change in the square root $\sqrt{\vbr^2}$, with a sudden change in the sign of the imaginary part $\Im\left(\sqrt{\vbr^2}\right)$. If one integrates across this discontinuity, the dependence on $t$ of the Coulomb integral \eqref{Coulomb-correction-phase} ceases to be analytic, and one loses the freedom that allowed the real contour of \eqref{initial-temporal-integral} to be deformed to pass through the saddle point $\ts$ in the first place, and the results are no longer meaningful.

The solution to this problem is to view the entire potential as a function of time,
\begin{equation}
U(\rcl(t)) = -\frac{1}{\sqrt{\rcl(t)^2}},
\end{equation}
as a single analytical function of the complex-valued variable $t$. The branch cuts in $U$ are imprinted on the time plane via the conformal mapping $t\mapsto \rcl(t)^2$. We show in \reffig{fig:complex-position-contour-plot} an example of the Coulomb potential's behaviour, as a contour map of the function $\sqrt{ \rcl(t)^2 }$ (which has the branch cut structure of $U$ but omits its singularities). The essential features of this function are the branch cuts, which are sketched in \reffig{fig:branch-cut-sketch} in red. In particular, the standard integration contour of \reffig{fig:standard-time-contour} -- straight down from $\ts$ and then along the real axis, shown dotted in \reffig{fig:branch-cut-sketch} -- can indeed cross the branch cuts when the electron returns near the ion.

This means that to preserve the analyticity of \eqref{Coulomb-correction-phase} one must deform the integration contour away from the real time axis until the integrand is continuous and analytic. This will correspondingly change the way the complex position $\rcl(t)$ moves in space, with the main effect of \textit{minimizing} the imaginary part of the position at the time of recollision; this is exemplified in \reffig{fig:complex-z-curved-contours}.

The relationship between the chosen temporal contour and the corresponding trajectory in complex position space, particularly along $z$, is in general complicated and hard to visualize. Similarly, the final momentum $\vbp$ has a strong effect on the potential $U(\rcl(t))$ as a function of time, sometimes very sensitively (such as near a soft recollision). These variables are all intertwined, with the momentum determining the branch cut structure and therefore the possible contours on both the time and space complex planes.

To help disentangle these relationships, we have developed a software package \cite{QuantumOrbitsDynamicDashboard} which enables real-time visualization of the effect of a particular choice of contour and momentum on the space trajectory and on the temporal branch cut structure. This software is described in detail in \citer{QODD_Software_paper} and we encourage the reader to use its visualization tools to explore the implications of our results.

\section{Times of closest approach}
\label{sec:times-of-closest-approach}
We see, then, that the existence of branch cuts which can cross the real axis can preclude the use of the standard integration contour for the integral in \eqref{Coulomb-correction-phase}, and one needs to choose a contour which passes through the `slalom gate' left by the branch cuts of $U(\rcl(t))$ as in \reffig{fig:complex-position-contour-plot}. If only a few momenta are involved then this can be done on a case-by-case basis, but the computation of photoelectron spectra requires a more programmatic approach, which is able to automatically choose the correct contour for each given momentum.

\begin{figure}[b]
  \begin{tabular}{rl}
  \includegraphics[width=0.6\columnwidth]{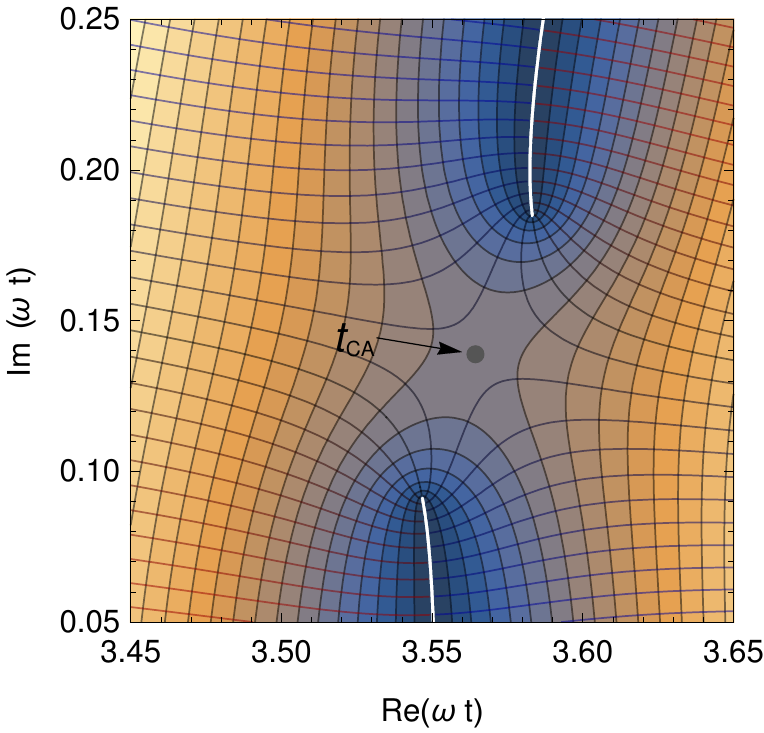}
  &
  \hspace{-5mm}
  \rotatebox{90}{
    \begin{tabular}{ccc}$\,\,\,\,\,\,\,\,$&
      \rotatebox{-90}{\includegraphics[scale=1]{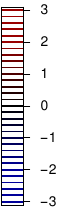}}
      &
      \hspace{-2mm}
      \rotatebox{-90}{\includegraphics[scale=1]{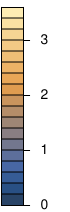}}
      \vspace{1mm}\\\hspace{21pt}& $\scriptscriptstyle \Im \left(\sqrt{\rcl(t)^2}\right)$ & $\scriptscriptstyle \Re\left(\sqrt{\rcl(t)^2}\right)$
    \end{tabular}
  }
  \end{tabular}
  \caption{
  Closer view of the region between the branch cuts in \reffig{fig:complex-position-contour-plot}. The space between any two branch cuts always contains a saddle point $\tca$, as this is the only way to join the curvatures of the contours next to each branch cut. For contours that pass through it, the saddle point marks the minimum of $\Re\left(\sqrt{\rcl(t)^2}\right)$. For valid contours that do not pass through $\tca$, this minimum will be smaller.
  }
  \label{fig:tCA-zoom}
\end{figure}

To obtain this approach, we examine in detail the space between the two branch cuts, as shown in \reffig{fig:tCA-zoom}. Each branch cut is a contour of constant $\Re(\sqrt{\rcl(t)^2})=0$, which means that the neighbouring contours must closely follow its direction, and circle around it when it terminates at the branch point. In the `slalom gate' configuration of \reffig{fig:complex-position-contour-plot}, the only way to combine the curvatures of the contours near the facing branch cuts is to have a saddle point in between them. This saddle point is the crucial object which enables the automatic choice of a contour which avoids the branch cuts, as it is ideally situated between them. Moreover, it has a deep physical and geometrical significance, which we explore in this section.

It is clear in \reffig{fig:tCA-zoom} that any contour which crosses the complex $t$ plane from left to right must have a real distance to the origin $\Re(\sqrt{\rcl(t)^2})$ which decreases, reaches a minimum, and then increases again. For contours which cross the branch cut, this minimum is zero, and it is accompanied by a discontinuous change in the imaginary part. For a path crossing through the saddle point, on the other hand, the minimum of $\Re(\sqrt{\rcl(t)^2 })$ is at its maximum value, and this minimum occurs precisely at the saddle point. 

For this reason, we call the saddle point the time of closest approach, and we label it $\tca$. To be precise, then, a contour that passes through $t=\tca$ maximizes the minimum value of the real part of the distance to the origin, $\Re(\sqrt{\rcl(t)^2})$. Intuitively, it permits the furthest possible approach to the ion.

Similarly, when calculating the Coulomb potential along such a contour, this choice minimizes the maximum value of the real part of $1/\sqrt{\rcl(t)^2}$ and its absolute value (for valid contours which do not cross the branch cuts), so that the Coulomb interaction is kept as bounded as possible. As long as the potential $U(\rcl(t))$ stays continuous and analytic, this is not essential, as the integral in \eqref{Coulomb-correction-phase} does not change. However, this contour optimizes the applicability of the approximations that led to \eqref{Coulomb-correction-phase}, and it admits the clearest physical interpretation by keeping the imaginary part of the position within the tightest bound possible at the points where this is necessary. Additionally, it minimizes the difficulties faced by numerical integration routines.

To find the saddle points, one simply looks for zeros of the derivative of $\sqrt{\rcl(t)^2 }$. This can be further reduced to the zeros of $\tfrac{\d}{\d t}\left[\rcl(t)^2\right]$, so the criterion is simply
\begin{equation}
\rcl(\tca)\cdot\vbv(\tca)=0.
\label{tca-equation}
\end{equation}
This equation is deceptively simple, and one must remember that the left-hand side is a complex-valued function of time through Eq.~\eqref{complex-trajectory}. Nevertheless, it has a compelling physical interpretation, for if a classical electron passes near the nucleus then it is closest to the origin when its velocity and its position vector are orthogonal. 

In this spirit, then, it is worthwhile to investigate the classical solutions of \eqref{tca-equation} before exploring the solutions in the complex quantum domain. As we shall see, both domains exhibit rich geometrical structures which are closely related to each other and, moreover, the times and momenta of soft recollisions emerge naturally as crucial geometrical points within both structures. After exploring the geometrical implications in both contexts, we shall use this knowledge to automatically generate correct integration contours for any momentum.

\subsection{Classical times of closest approach}
The classical closest-approach times $\tcacl$ are naturally defined via
\begin{equation}
\Re\left[\rcl(\tcacl)\cdot\vbv(\tcacl)\right]=\Re\left[\rcl(\tcacl)\right]\cdot\vbv(\tcacl)=0.
\label{tca-classical-equation}
\end{equation}
The problem in using them as a tool, however, is that both of their defining equations, \eqref{tca-equation} and \eqref{tca-classical-equation}, have several other solutions, and these are not always distinguishable from the desired closest-approach times. In particular, the turning points of the classical trajectory, away from the core, are also solutions of \eqref{tca-equation}, since they are also extrema of $\vbr^2$. Thus, for instance, the surface in \reffig{fig:complex-position-contour-plot} contains saddle points at $\omega t\approx \pi/4$ and $\omega t\approx 3\pi/4$, which correspond to the turning points shown in \reffig{fig:complex-z-curved-contours} to the right and left of the position at $\tn$, respectively. Thus, to be able to use the closest-approach times as an effective tool to avoid the branch cuts one needs to distinguish the crucial mid-gate $\tca$ points from the other solutions.

\begin{figure}[b]
\begin{tabular}{cc}
\raisebox{25mm}{\subfigure[]{\makebox[5mm][0mm]{} }}\hspace{-11mm} &
\addtocounter{subfigure}{-1}
\subfigure{
  \includegraphics[width=0.9\columnwidth]{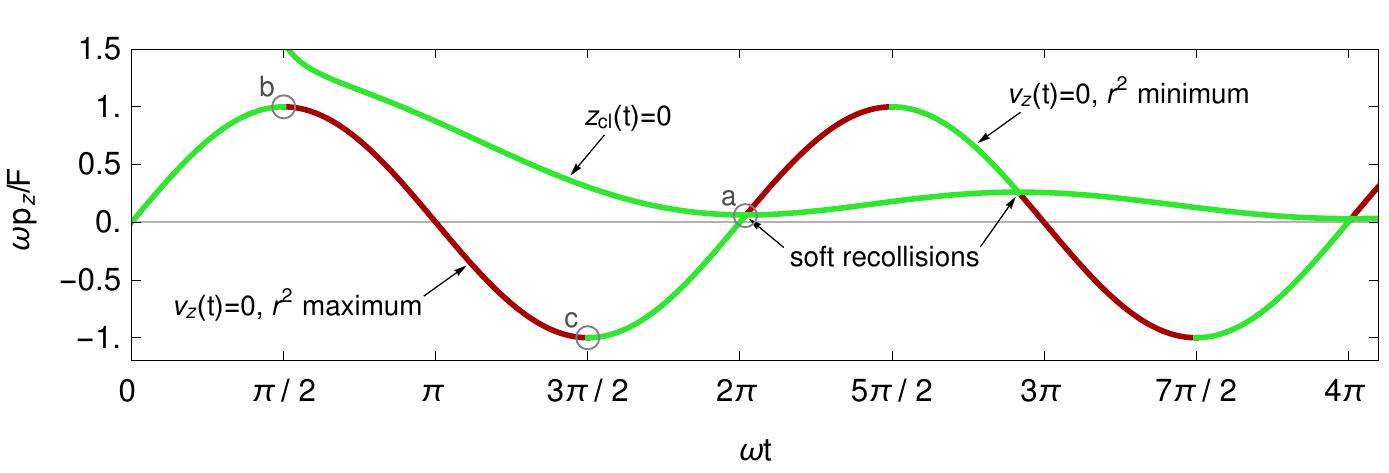}
  \label{fig:classical-tca-on-axis}
  }
  \\
\raisebox{40mm}{\subfigure[]{\makebox[5mm][0mm]{} }}\hspace{-4mm} &
\addtocounter{subfigure}{-1}
\subfigure{
  \includegraphics[width=0.9\columnwidth]{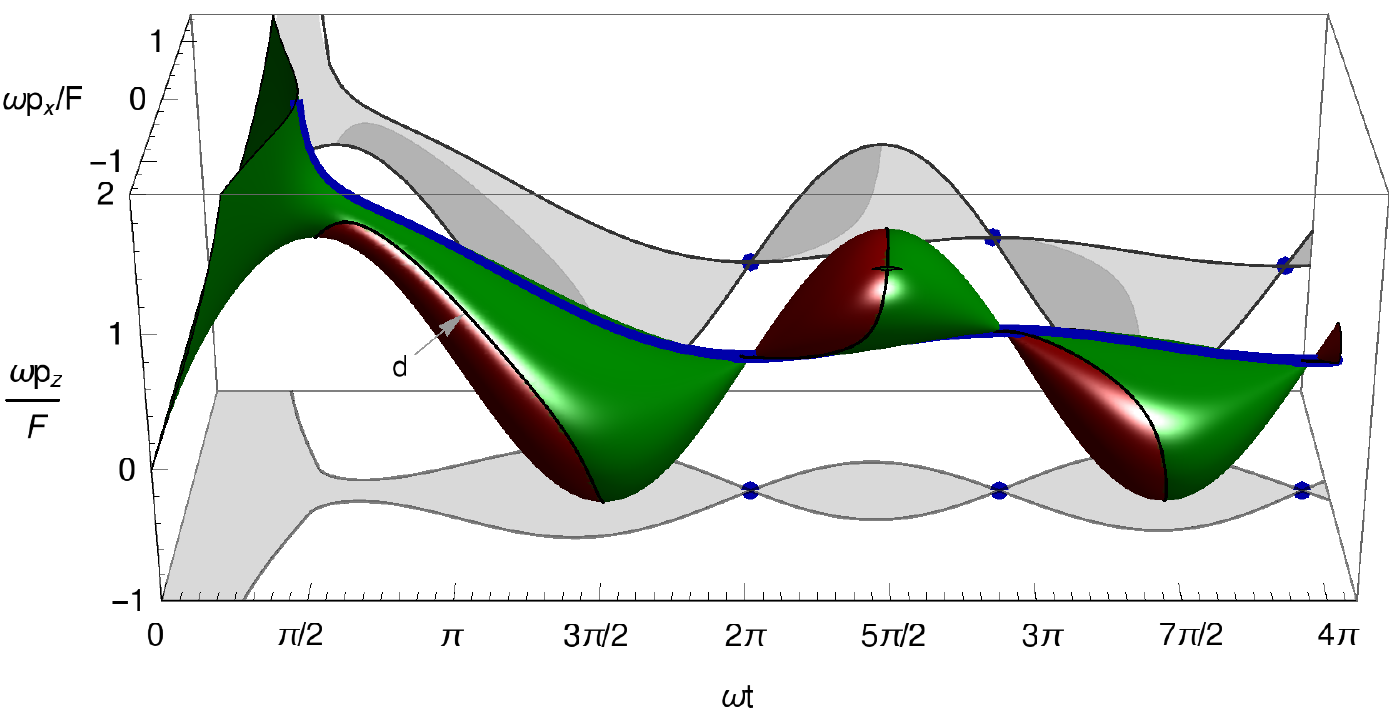}
  \label{fig:classical-tca-surface}
  }
\end{tabular}
  \caption{ 
  The classical closest-approach times with $p_\perp=0$, satisfying \eqref{tca-classical-equation-on-axis}, separate into two curves  \subref{fig:classical-tca-on-axis}: turning points with $v_z=0$ and collisions with $\zcl=0$, with their intersections representing soft recollisions. At the points marked \texttt{a}, \texttt{b} and \texttt{c} the different roots merge and disappear, with the corresponding trajectories displayed in \reffig{fig:sample-trajectories}.
  For nonzero $p_\perp$, the solutions of the vector equation \eqref{tca-classical-equation} form a single coherent surface \subref{fig:classical-tca-surface} with a sequence of bounded lobes which connect at the soft recollisions, where the surface is locally a cone.
  Local minima and maxima of $\rcl(t)^2$ are shown respectively in green and red in both panels. At the boundary between the two, a maximum and minimum meet, merge and disappear, as shown in the trajectory of \reffig{fig:sample-trajectories-d}; the point then leaves the surface.
  The side  and bottom panels show the projections of the surface on $p_z$ and $p_x$ respectively. 
  An interactive 3D version of this figure is available in the Supplementary Information \cite{SupplementaryInformation}.
 }
  \label{fig:classical-tca-plots}
\end{figure}

In certain cases this is easy, such as for on-axis collisions with $p_\perp=0$. Here \eqref{tca-classical-equation} reads
\begin{equation}
\Re\left[\zcl(\tcacl)\right]v_z(\tcacl)=0,
\label{tca-classical-equation-on-axis}
\end{equation}
and it solutions cleanly separate into turning points, with $v_z(\tcacl)=0$, and collisions with $\zcl(t)=0$, as shown in \reffig{fig:classical-tca-on-axis}. The turning points can additionally be classified as minima and maxima of $\rcl(t)^2$, shown respectively in green and red, by evaluating the sign of $\tfrac{\d^2}{\d t^2}\rcl(t)^2$. At nonzero $p_\perp$, it is the collisions that will turn into useful closest-approach times.

Within this context, the classical soft recollisions studied in Section~\ref{sec:classical-transitions} appear naturally as the intersection between these two curves: the times when both $v_z(t)=0$ and $\zcl(t)=0$. Moreover, at these intersections the number of available roots changes. Thus, as $p_z$ is swept up across the intersection marked \texttt{a} in \reffig{fig:classical-tca-on-axis}, an inward turning point turns into an outward turning point flanked by two closest-approach points. The classical trajectory exhibits exactly this behaviour, as shown in \reffig{fig:sample-trajectories-a}, with this change in $p_z$.

The roots of \eqref{tca-classical-equation-on-axis} can also merge at the extremes of the sinusoidal turning-point curve of \reffig{fig:classical-tca-on-axis}. At these points, the longitudinal momentum becomes greater than the oscillation amplitude $F/\omega$, and the velocity $v_z(t)=p_z-\tfrac F\omega \sin(\omega t)$ no longer changes sign. The resulting behaviour of the trajectories is shown in Figs.~\ref{fig:sample-trajectories-b} and \subref{fig:sample-trajectories-c}, and resembles pulling a winding string until the turns are straight.

The closest-approach solutions of \eqref{tca-classical-equation} become more interesting when one allows a nonzero transverse momentum~$p_x$. Here the solutions form a single coherent surface, shown in \reffig{fig:classical-tca-surface}, that consists of a number of bounded lobes joined together at the soft recollisions, which locally look like cones. Thus, it is possible to continuously connect any two roots of \eqref{tca-classical-equation} via a path on the surface. More specifically, the inward turning points and the recollisions shown in green in \reffig{fig:classical-tca-on-axis} can be smoothly connected via the $p_x\neq0$ component of the surface, which precludes the existence of a simple criterion to distinguish one from the other in the general case.

\newlength{\figsampletrajectoriesheight}
\setlength{\figsampletrajectoriesheight}{0.465\columnwidth}
\begin{figure}[t]
  \begin{tabular}{cc}  
  \begin{tabular}[b]{cc}    
    \subfigure{\label{fig:sample-trajectories-a}}
    \includegraphics[height=\figsampletrajectoriesheight]{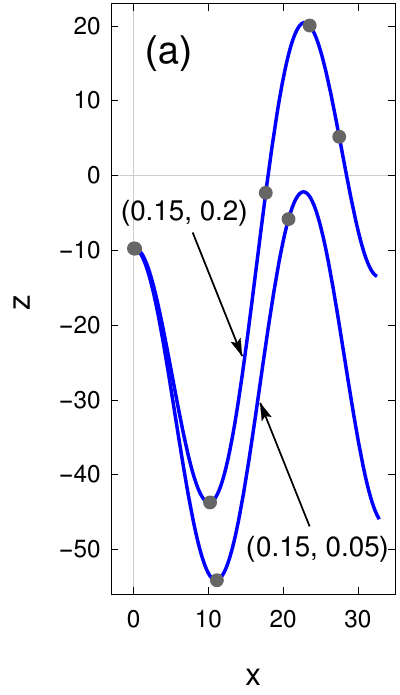}
    &
    \subfigure{\label{fig:sample-trajectories-b}}
    \includegraphics[height=\figsampletrajectoriesheight]{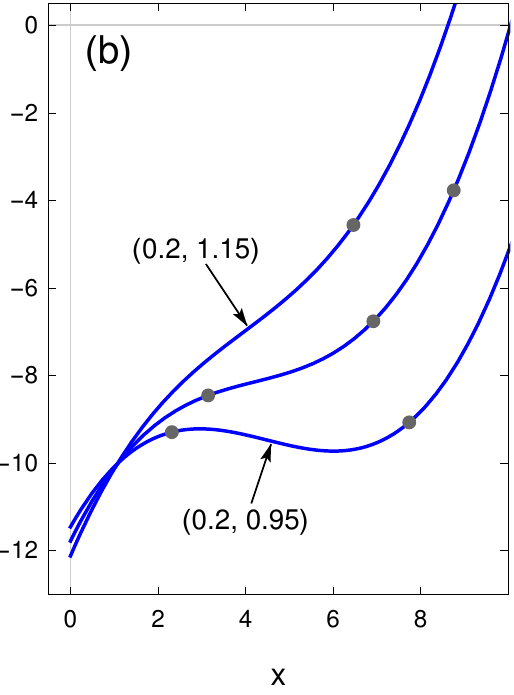} 
    \\
    \multicolumn{2}{c}{
    \subfigure{\label{fig:sample-trajectories-c}}
      \includegraphics[height=0.7\figsampletrajectoriesheight]{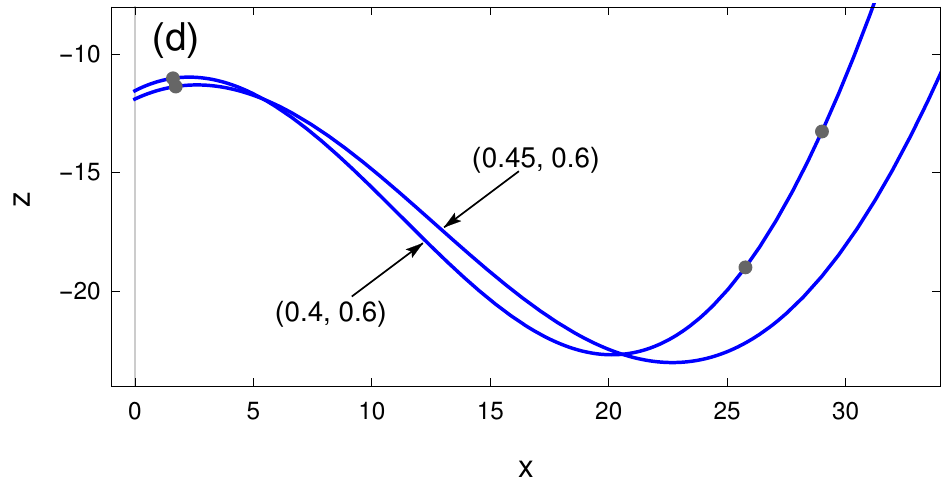}
    }  
  \end{tabular}
  &
    \subfigure{\label{fig:sample-trajectories-d}}
  \includegraphics[height=1.73\figsampletrajectoriesheight]{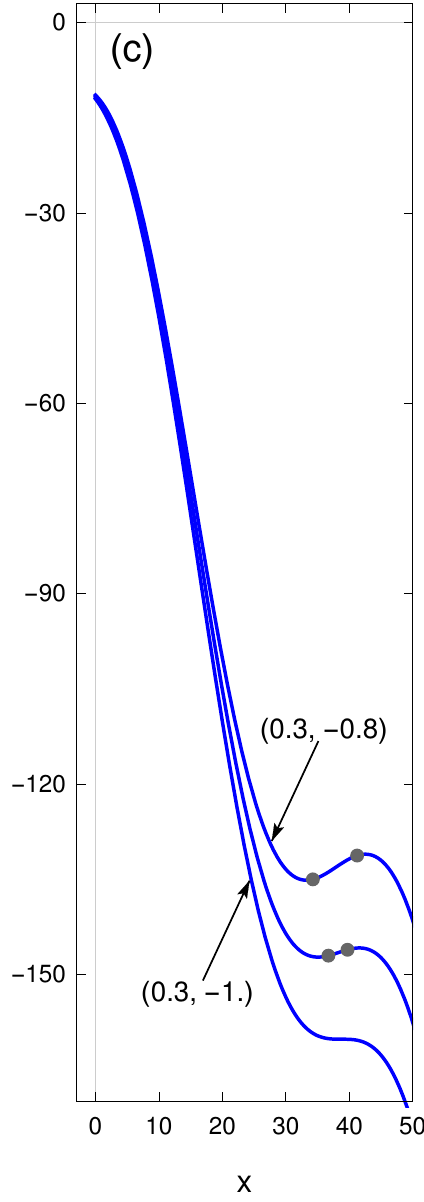}
  \end{tabular}
  \caption{
    Classical trajectories from near the critical points marked \texttt{a}-\texttt{d} in \reffig{fig:classical-tca-plots}, where the closest-approach roots of \eqref{tca-classical-equation} can merge and disappear, indexed by their reduced momentum $(\omega p_x/F,\omega p_z/F)$. The dots show closest-approach points, which satisfy \eqref{tca-classical-equation} and for which the tangent to the trajectory is orthogonal to the radius vector.
    In the neighbourhood of a soft recollision, \subref{fig:sample-trajectories-a}, an inward turning point turns into an outward turning point flanked by closest-approach points as the momentum increases.
    In \subref{fig:sample-trajectories-b} and \subref{fig:sample-trajectories-c} two turning points, a maximum and a minimum of $\rcl(t)^2$, merge and disappear as $p_z$ goes past the oscillation amplitude $F/\omega$, either in the positive \subref{fig:sample-trajectories-b} or negative \subref{fig:sample-trajectories-c} direction, so $v_z$ no longer changes sign and no turning points occur.
    Similarly, as the transverse momentum $p_x$ increases in \subref{fig:sample-trajectories-d} the $x$ component of the velocity becomes too great for the tangent to be orthogonal to the radius vector; at that point a maximum and minimum of $\rcl(t)^2$ merge and disappear, leaving $\rcl(t)^2$ to grow monotonically.
    }
  \label{fig:sample-trajectories}
\end{figure}

On the other hand, the outward turning points can still be distinguished, as they are local maxima of $\rcl(t)^2$. These maxima are shown in red in Figs.~\ref{fig:classical-tca-on-axis} and \subref{fig:classical-tca-surface}, and they form the ``left-facing'' side of the surface in \reffig{fig:classical-tca-surface}. More specifically, any horizontal line of constant momentum must enter the surface through a maximum (red) and leave it through a minimum (green), because the minima and maxima must alternate for any given trajectory. Thus, the red (maximum) side of the surface points towards negative $t$, and the green (minimum) side points towards positive $t$.

At the boundary between both parts of the surface, a maximum and a minimum merge and disappear, and the trajectory will then behave as shown in Fig.~\ref{fig:sample-trajectories-b}, \subref{fig:sample-trajectories-c} or \subref{fig:sample-trajectories-d}, depending on which direction the boundary is approached (i.e. towards positive $p_z$, negative $p_z$, and increasing $p_x$, respectively). Horizontal lines of constant momentum will be tangent to the surface at this boundary, and the corresponding trajectory will have a double root of \eqref{tca-classical-equation}.

\subsection{The quantum solutions}
The quantum solutions have a richer geometry, with an additional dimension, imaginary time, to be occupied. The main effect of this is to increase the number of available solutions: while in the classical case two real solutions of \eqref{tca-classical-equation} can merge and disappear, in the quantum case the complex solutions of \eqref{tca-equation} are not lost, but will move into imaginary time and remain present.

\begin{figure}[b]
  \includegraphics[width=\columnwidth]{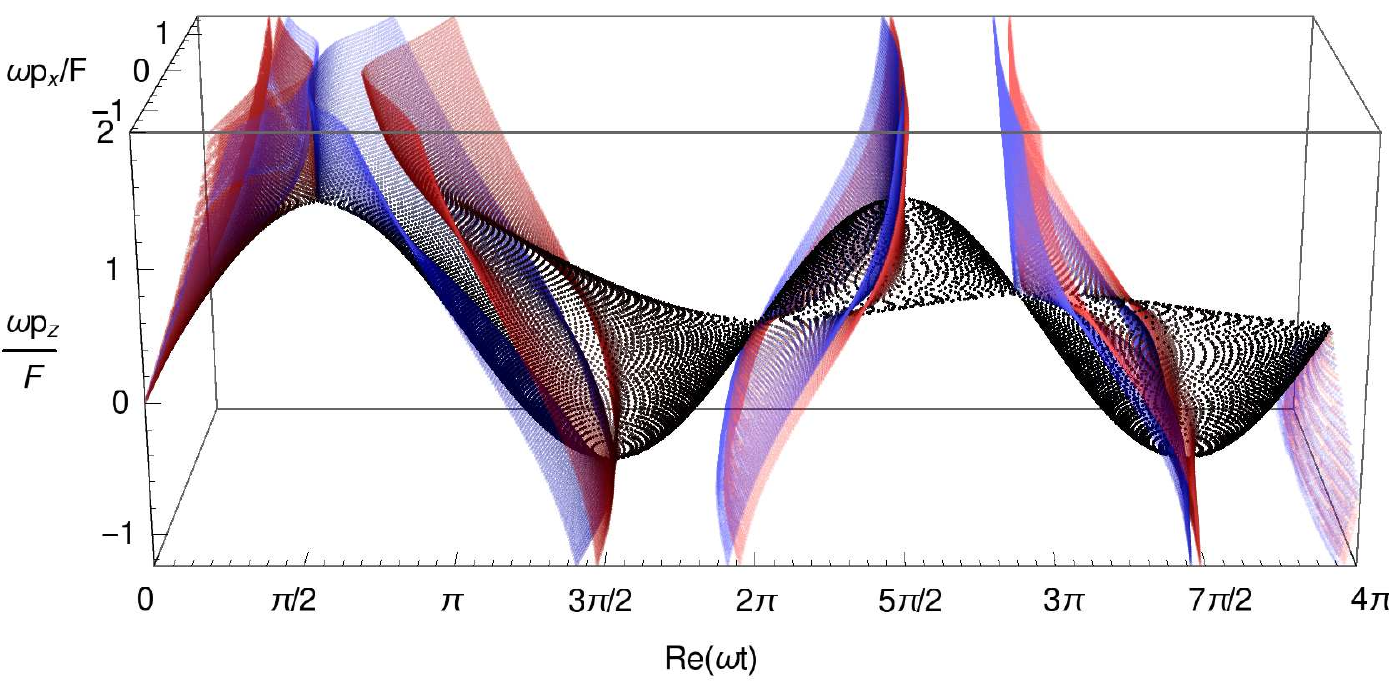}
  \caption{ 
  The quantum solutions of the closest-approach equation \eqref{tca-equation} form multiple surfaces which wrap around the classical solutions of \reffig{fig:classical-tca-surface}, closely following the lobes where they exist and departing at the edges to form pairs of parallel surfaces with imaginary parts of opposite sign. 
  Black dots represent largely real solutions, with red (blue) dots representing solutions with positive (negative) imaginary part.
  An interactive 3D version of this figure is available in the Supplementary Information \cite{SupplementaryInformation}.
 }
  \label{fig:quantum-tca-surface}
\end{figure}

In general, the quantum solutions will be close to the classical ones when the latter exist. As one approaches the end of a lobe on the classical solution surface of \reffig{fig:classical-tca-surface}, however, the quantum solutions approach each other close to the real axis and then diverge into positive and negative imaginary time, keeping a relatively constant real part. If one then projects this to real times, the result is a pair of surfaces which closely follow the red and green parts of the classical surface, and then diverge into roughly parallel planes as they reach the boundary. We show this behaviour in~\reffig{fig:quantum-tca-surface}.

The first few closest-approach solutions are relatively easy to handle, and depend smoothly on the momentum. This includes the first minimum and maximum of $\rcl(t)^2$, like the ones in \reffig{fig:complex-position-contour-plot}, the birth time $\ts$ itself, and a conjugate solution with negative imaginary part which should be ignored. These solutions occupy specific regions of the complex $t$ plane, as shown in \reffig{fig:tca-grid-patterns}, and can be consistently identified. Moreover, these solutions exhibit close approaches at $\omega t\approx \pi/2$ and $\omega t\approx 3\pi/2$ which are the quantum counterparts of the classical maximum-minimum mergers shown in Figs.~\ref{fig:sample-trajectories-b} and~\subref{fig:sample-trajectories-c}. These are evident in \reffig{fig:quantum-tca-surface} as the converging surfaces at those times, and are of relatively limited interest.

The most important $\tca$ close approaches occur at and near the soft recollisions, shown inside the gray rectangle  of \reffig{fig:tca-grid-patterns} and in \reffig{fig:tca-mixing-paths}, with a complicated momentum dependence which we explore below. In the quantum domain, soft recollisions again represent interactions between three different closest-approach roots. Unlike the classical domain, however, the roots do not merge; instead, two of them move into imaginary time after a three-way avoided collision, shown in \reffig{fig:tca-mixing-paths} between the points marked 9 and 10. The proximity between the multiple saddle points marks the increased time the electron spends near the ion in the neighbourhood of a soft recollision.

\newlength{\figurenineheight}
\setlength{\figurenineheight}{0.592 \columnwidth}
\begin{figure}[t]
\centering
\begin{tabular}{c}
\raisebox{41mm}{\subfigure[]{\makebox[5mm][0mm]{} }}\hspace{-7.5mm} 
\addtocounter{subfigure}{-1}
\subfigure{
  \includegraphics[width=0.95\columnwidth]{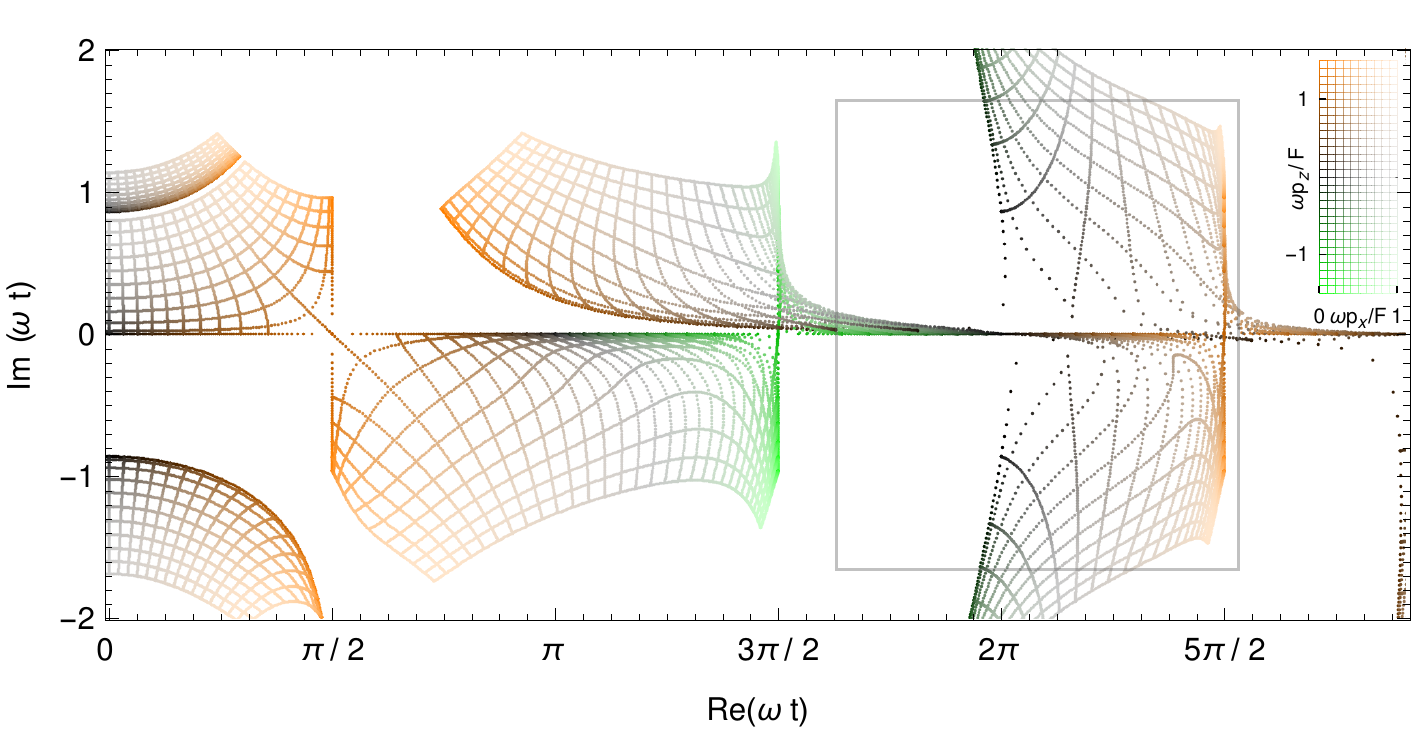}
  \label{fig:tca-grid-patterns}
}
\\
\begin{tabular}{cc}
\raisebox{52mm}{\subfigure[]{\makebox[5mm][0mm]{} }}\hspace{-8.5mm} 
\addtocounter{subfigure}{-1}
\subfigure{
  \includegraphics[height=\figurenineheight]{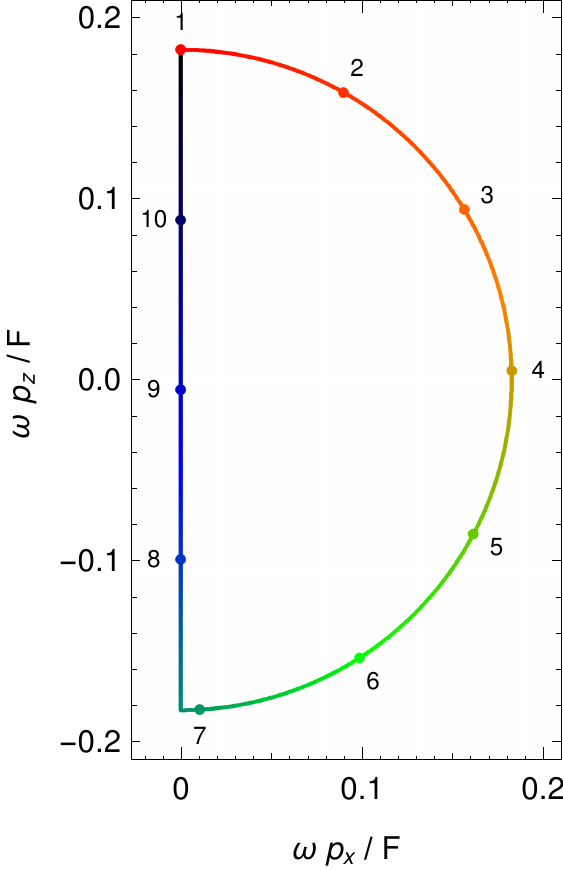}
  \label{fig:momentum-semicircle}
  }
\raisebox{52mm}{\subfigure[]{\makebox[5mm][0mm]{}}}\hspace{-8.5mm} 
\addtocounter{subfigure}{-1}
&
\subfigure{
  \includegraphics[height=\figurenineheight]{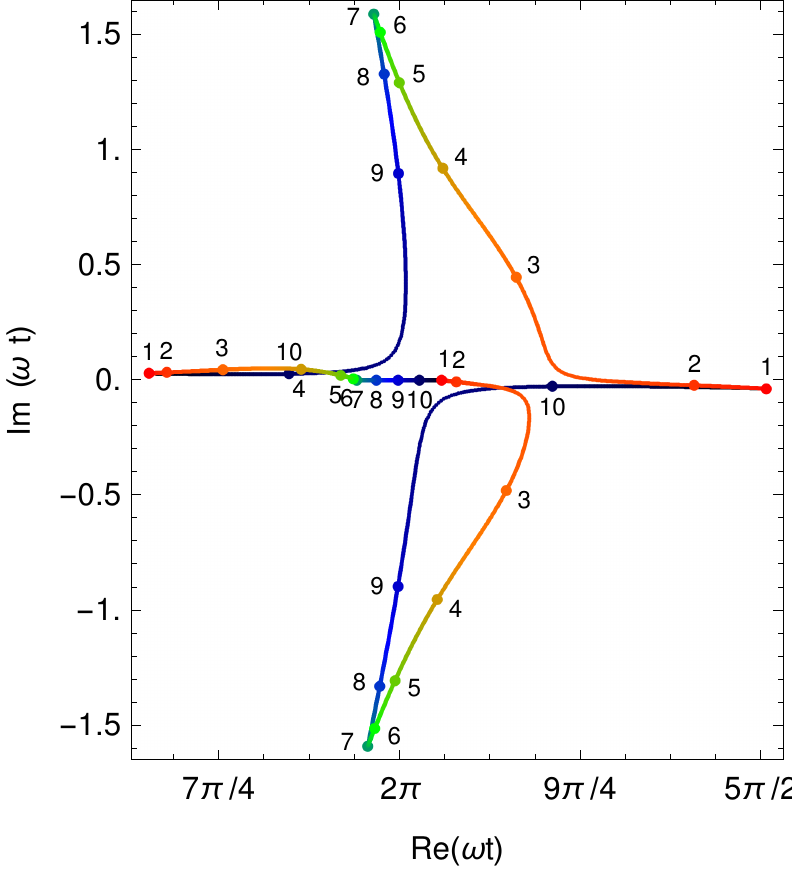}
  \label{fig:tca-mixing-paths}
  }
\end{tabular}
\end{tabular}
  \caption{ 
  The quantum closest-approach times, which satisfy \eqref{tca-equation}, on the complex time plane. In \subref{fig:tca-grid-patterns} we show the closest approach times corresponding to a grid in momentum space (shown inset). These are generally grid-like in the time plane, though after $\omega t=3\pi/2$ some solutions lie close to the real axis and cannot be discerned in this view.
  At the cusps of Figs.~\ref{fig:classical-tca-surface} and \ref{fig:quantum-tca-surface}, which correspond to soft recollisions, the regular grid-like behaviour breaks and the solutions can no longer be uniquely tagged. 
  This can be seen by following the semicircular path in momentum space shown in \subref{fig:momentum-semicircle}, for which there are three solutions in the gray rectangle of \subref{fig:tca-grid-patterns}, shown in detail in \subref{fig:tca-mixing-paths}. Going once around the semicircle moves the $\tca$ along the bow-shaped curves and, upon returning to the initial point, permutes them cyclically.
  In particular, this path contains an avoided collision between the points marked 2 and 3, and a three-way avoided collision between points 9 and 10. This three-way interaction marks the soft recollision itself.
  An interactive 3D version of this figure, with considerable additional detail, is available in the Supplementary Information \cite{SupplementaryInformation}.
 }
  \label{fig:complex-tca-plane-structures}
\end{figure}

More interestingly, this three-way collision marks a crucial topological change in the configuration of the branch cuts associated with the recollision, as shown in \reffig{fig:branch-cut-topology-change}. Each of the outer saddle points, $\tcasup{\,(1)}$ and $\tcasup{\,(3)}$, has a pair of branch cuts associated with it, in the same `slalom gate' configuration as in \reffig{fig:tCA-zoom}, and these go off into imaginary time. However, the way in which they do so changes as the longitudinal momentum $p_z$ passes the momentum $\pzsr$ of the soft recollision.

For $p_z$ below $\pzsr$, as in \reffig{fig:branch-cut-topology-open}, the branch cuts loop back to imaginary time without crossing the real time axis. As with the low-momentum trajectory of \reffig{fig:sample-trajectories-a}, the trajectory does not quite reach the collision, and the associated branch cuts do not force a change of contour. At $p_z=\pzsr$, however, the branch cuts touch and reconnect, and for $p_z>\pzsr$ the topology changes to the one shown in \reffig{fig:branch-cut-topology-closed}. Here the trajectory does pass the core, and the associated branch cuts do cross the real axis, forcing the integration contour to change and pass through the gates.

This process has profound implications for the ionization amplitude, because these drastic changes in the integrand occur precisely when it is largest. Thus, choosing the wrong contour in this region accounts for the largest contributions to the integrand, with a correspondingly large effect on the integral. More surprisingly, once the contour is forced to pass through the `gate' $\tca$s, for $p_z$ just above $\pzsr$, their contributions have the effect of suppressing the ionization amplitude there.

To see how this comes about, consider the integral $\int U(\rcl(t))\d t$ for the configuration of \reffig{fig:branch-cut-topology-open}. Here $\sqrt{\rcl(t)^2}$ has a minimum at the central saddle point, $\tcasup{\,(2)}$, and this translates into a maximum of $1/\sqrt{\rcl(t)^2}$ which dominates the integral. At this point, the approach distance $r_\ast=\sqrt{\rcl(\tcasup{\,(2)})^2}$ is dominated by a modest and positive imaginary part. This means that $U_\ast=-1/r_\ast$ is large and imaginary, and therefore the correction factor $e^{-i\int U\d t}$ has a large amplitude.

On the other hand, in the configuration of \reffig{fig:branch-cut-topology-closed} the integral is dominated by the `gate' closest-approach times, for  which $r_\ast'=\sqrt{\rcl(\tcasup{\,(1)})^2}\approx\sqrt{\rcl(\tcasup{\,(1)})^2}$ is mostly real and much smaller than $r_\ast$. The corresponding potential $U_\ast'=-1/r_\ast'$ is then large, real and negative, and $-iU_\ast'$ is along $+i$. However, here the line element $\d t$ must slope upwards with a positive imaginary part to emphasize the contribution of the saddle point, and this then gives $-i\int U(\rcl(t))\d t$ a large and negative real part. This, in turn, suppresses the amplitude of the correction factor $e^{-i\int U\d t}$.

This effect is visible in photoelectron spectra as a large peak just below the soft recollision, followed by a deep, narrow dip. In an experimental setting, the dip will almost certainly get washed out by nearby contributions unless specific steps are taken to prevent this, but the peak will remain. In addition, this effect mirrors the redistribution of population seen in classical-trajectory based approaches, where the peaks caused by dynamical focusing represent trajectories taken from other asymptotic momenta, whose amplitude is reduced.

\newcommand{\figuretenscale}{0.89}
\begin{figure}[!t]
  \begin{tabular}{ccl}
  $\qquad p_z=\figuretwelveppl{}F/\omega$   &    $p_z=\figuretwelvepph{}F/\omega$   &
  \\  \hline  \vspace{-2mm} \\
  {\scriptsize\hspace{6mm}(a) $\sqrt{\rcl(t)^2}$  }
  &
  {\scriptsize(b) $\sqrt{\rcl(t)^2}$}
  &  \vspace{-2.5mm}  \\
  \subfigure{
    \includegraphics[scale=\figuretenscale]{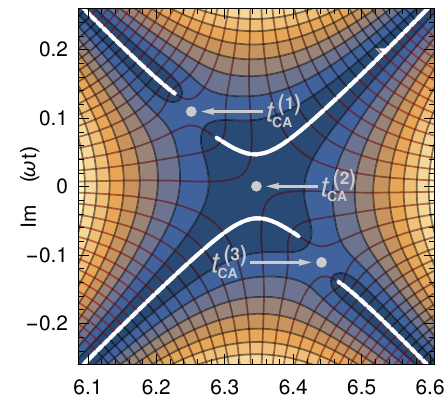}
    \label{fig:branch-cut-topology-open}
  }
  & \hspace{-3mm}
  \subfigure{
    \includegraphics[scale=\figuretenscale]{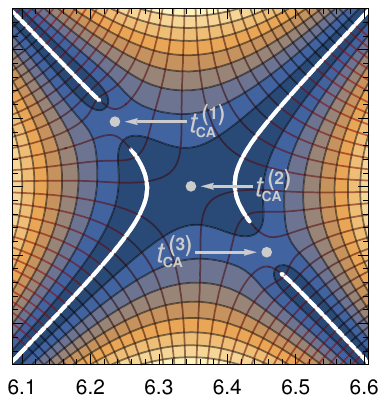}
    \label{fig:branch-cut-topology-closed}
  }
  &
  \hspace{-5mm}
  \rotatebox{90}{
    \begin{tabular}{ccc}\hspace{7pt}&
      \rotatebox{-90}{\includegraphics[scale=\figuretenscale]{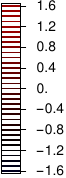}}
      &
      \hspace{-2mm}
      \rotatebox{-90}{\includegraphics[scale=\figuretenscale]{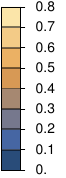}}
      \vspace{-1mm}\\ & $\scriptscriptstyle\Im$ & $\scriptscriptstyle\Re$
    \end{tabular}
  }
  \\
  {\scriptsize\hspace{6mm}(c) $\vbv(t)^2$}
  &
  {\scriptsize(d) $\vbv(t)^2$}
  &  \vspace{-2.5mm}  \\
  \subfigure{
    \includegraphics[scale=\figuretenscale]{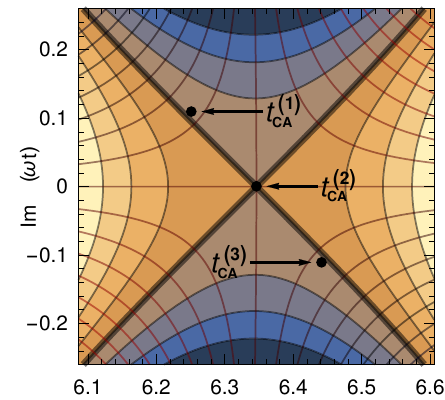}
    \label{fig:branch-cut-velocity-open}
  }
  & \hspace{-3mm}
  \subfigure{
    \includegraphics[scale=\figuretenscale]{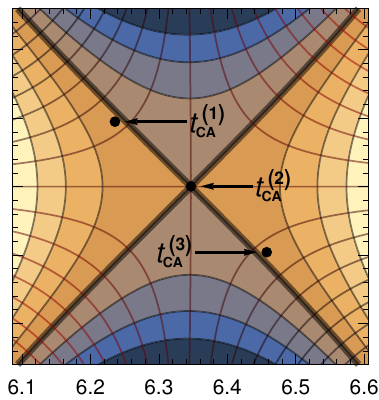}
    \label{fig:branch-cut-velocity-closed}
  }
  &
  \hspace{-5mm}
  \rotatebox{90}{
    \begin{tabular}{ccc}\hspace{7pt}&
      \rotatebox{-90}{\includegraphics[scale=\figuretenscale]{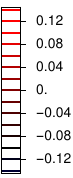}}
      &
      \hspace{-2mm}
      \rotatebox{-90}{\includegraphics[scale=\figuretenscale]{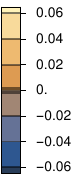}}
      \vspace{-1mm}\\\hspace{21pt}& $\scriptscriptstyle\Im$ & $\scriptscriptstyle\Re$
    \end{tabular}
  }
  \\
  {\scriptsize\hspace{6mm}(e) Branch cut sketch}
  &
  {\scriptsize(f) Branch cut sketch}
  &  \vspace{-2.5mm}  \\
  \subfigure{
    \includegraphics[scale=\figuretenscale]{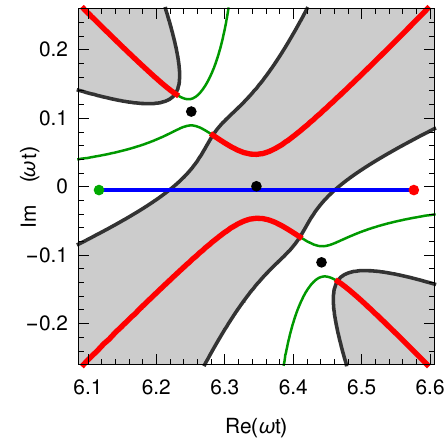}
    \label{fig:branch-cut-sketch-open}
  }
  & \hspace{-3mm}
  \subfigure{
    \includegraphics[scale=\figuretenscale]{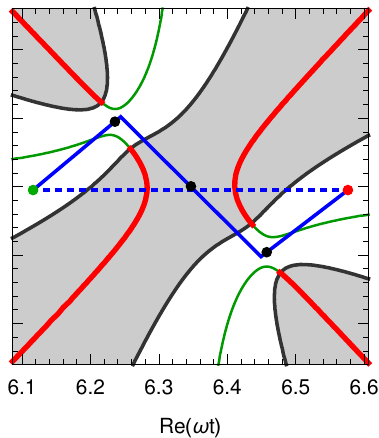}
    \label{fig:branch-cut-sketch-closed}
  }
  &
  \end{tabular}
  \begin{tabular}{cc}
  {\scriptsize\hspace{6mm}(g) Trajectory}
  &
  {\scriptsize(h) Trajectory}
  \vspace{-2.5mm}  \\
  \subfigure{
    \includegraphics[scale=\figuretenscale]{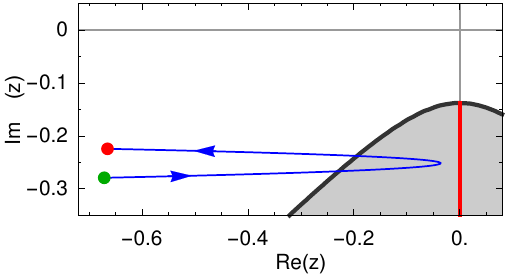}
    \label{fig:branch-cut-trajectory-open}
  }
  & \hspace{-4mm}
  \subfigure{
    \includegraphics[scale=\figuretenscale]{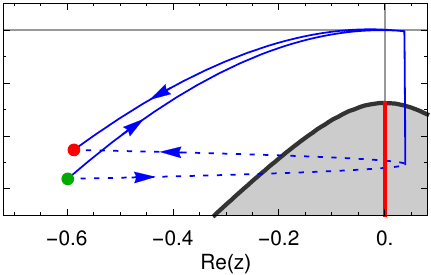}
    \label{fig:branch-cut-trajectory-closed}
  }
  \end{tabular}
  \vspace{0mm}
  \caption{ 
    During a soft recollision, the quantum times of closest approach will perform a three-way close approach, like that shown in \reffig{fig:tca-mixing-paths} at the first soft recollision between the points 9 and 10. At this close approach, the branch cuts associated with these saddle points will reconnect and change topologies, as shown in \subref{fig:branch-cut-topology-open} and \subref{fig:branch-cut-topology-closed}. 
    This happens near, but not at, the classical soft-recollision momentum studied in Section~\ref{sec:classical-transitions}.
    At the point of the topological change, the outer saddlepoints $\tcasup{\,(1)}$ and $\tcasup{\,(3)}$ emerge from the classically forbidden region, and the real part of their kinetic energy $\Re(\tfrac12\vbv(t)^2)$ changes sign, as shown in \subref{fig:branch-cut-velocity-open} and \subref{fig:branch-cut-velocity-closed}. After the change, the outer saddlepoints do not require tunnelling to get to, and should be crossed by the integration contour.
    We show in \subref{fig:branch-cut-sketch-open} and \subref{fig:branch-cut-sketch-closed} a sketch of the relevant branch cuts, analogous to \reffig{fig:branch-cut-sketch}, along with the appropriate integration contour for each topology.
    In \subref{fig:branch-cut-trajectory-open} and \subref{fig:branch-cut-trajectory-closed} we show the corresponding trajectories in the complex $z$ plane for those contours. The motion along $z$ is similar to \reffig{fig:sample-trajectories-a}, with a modest imaginary part as in \reffig{fig:complex-z-trajectories}, and if the core is not reached does not represent a problem. For the slightly higher momentum of \subref{fig:branch-cut-trajectory-closed}, however, the trajectory passes the origin and would cross the associated branch cut if taken along the dashed integration contour of \subref{fig:branch-cut-sketch-closed}, so the contour must be deformed to avoid it, as shown by the solid line.
    Here $p_x=\figuretwelvepo$.
    \vspace{-8mm} 
  }
  \label{fig:branch-cut-topology-change}
\end{figure}

We now turn to the momentum dependence of the closest-approach times near the soft recollision, which again presents interesting topological features. The main problem is illustrated in Figs.~\ref{fig:momentum-semicircle} and \subref{fig:tca-mixing-paths}: the different solutions of \eqref{tca-equation} mix, and there is no longer any way to distinguish them from each other, as there was in the classical case. More specifically, traversing a closed loop in momentum space, like the semicircle shown in \reffig{fig:momentum-semicircle}, will move the roots around in such a way that when one returns to the initial point the overall configuration is the same, but the saddle points have been permuted cyclically.

Topologically, this means that the surface defined by \eqref{tca-equation} (a two-dimensional surface in a four-dimensional space) does not separate into distinct components; instead, the surface has a single connected component after $\omega t=3\pi/2$. On the other hand, the surface itself remains singly connected. Both of these behaviours are explored in Figs.~S3 and S4 in the Supplementary Information~\cite{SupplementaryInformation}.

This mixing behaviour is unusual in the quantum orbit formalism, where the norm is for rather elaborate indexing schemes to be possible \cite{Becker_rescattering, ATI_Saddle_point_classification}, partly because there is usually a single free parameter that governs the motion of the saddle points. Here the control space is two-dimensional, which allows for nontrivial closed loops inside it, and this defeats the possibility of attaching any type of label to individual roots of~\eqref{tca-equation}.

Finally, an interesting consequence of the mixing between roots is that, at certain specific values of $p_x$ and $p_z$, the roots must merge, giving double roots of~\eqref{tca-equation}. However, this behaviour depends very sensitively on the momentum, and it can safely be ignored. In fact, the very difficulty of tagging the roots, caused by the mixing, makes finding the merge momentum an elusive numerical problem.

\subsection{Navigating the branch cuts}

Apart from its noticeable effects on the ionization amplitude, the topological change in the temporal branch cuts has an obvious effect on the integration contour required to produce a correct yield. The closest-approach times are a useful tool in identifying and crossing the branch-cut gates when they are present. However, some gates lead to dead-end regions which do not need to be crossed, which means that not all $\tca$s are useful stepping points.

Thus, while the correct contour can always be chosen by hand given a specific branch cut configuration, it is also desirable to have an algorithm that will specify which `slalom gates' the contour should cross, and in which order it should do so. Without such an algorithm, it is impossible to automate the choice of contour and the calculation of photoelectron spectra is impractical.

Such an algorithm is indeed available, and it is based on a geometrical fact shown in Figs.~\ref{fig:branch-cut-velocity-open} and \subref{fig:branch-cut-velocity-closed}: the topological change in the branch cut connections that happens at a soft recollision comes together with a change in the sign of the real part of the squared velocity, $\vbv(t)^2$, for the outer saddle points. 

Thus, in the closed topology of \reffig{fig:branch-cut-topology-closed}, where the contour should pass through $\tcasup{\,(1)}$ and~$\tcasup{\,(3)}$, the real kinetic energy $\Re(\vbv(\tca)^2)$ is positive at those saddle points. In contrast, for the open topology of \reffig{fig:branch-cut-topology-open}, $\Re(\vbv(\tcasup{\,(1)})^2)$ and $\Re(\vbv(\tcasup{\,(3)})^2)$ are both negative, and the contour should ignore both of those saddle points.

The physical content of this criterion is quite clear: in the quantum orbit formalism, the classically forbidden regions are readily identified in the complex time plane as those regions where the kinetic energy $\tfrac12\vbv^2(t)$ is negative, or at least has negative real part. The undesirable saddle points of \reffig{fig:branch-cut-topology-open} therefore require the trajectory to tunnel towards the core to be reached. On the other hand, a formal proof of the simultaneity of the topological change with the emergence of the $\tca$ from the `barrier' is still lacking.

\begin{figure}[t]
  \begin{tabular}{c}
    (a) $\mathbf p=(\figurethirteenapo,0,\figurethirteenapp)$ \vspace{-3mm} \\
    \subfigure{
      \includegraphics[scale=1]{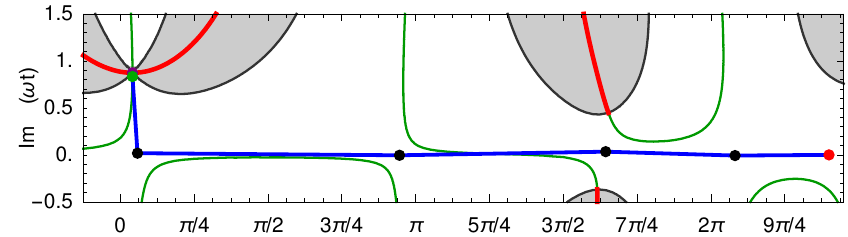}  
      \label{fig:path-chooser-examples-a}
    }
    \\
    (b) $\mathbf p=(\figurethirteenbpo,0,\figurethirteenbpp)$ \vspace{-3mm} \\
    \subfigure{
      \includegraphics[scale=1]{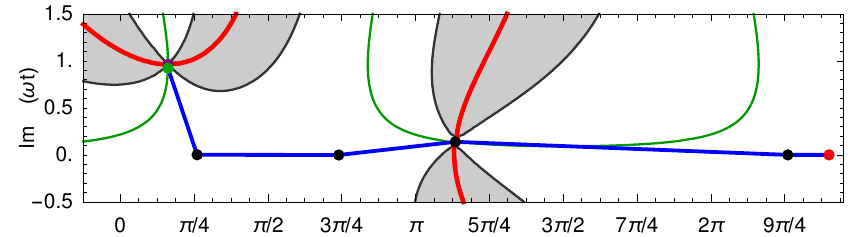}  
      \label{fig:path-chooser-examples-b}
    }
    \\
    (c) $\mathbf p=(\figurethirteencpo,0,\figurethirteencpp)$ \vspace{-3mm} \\
    \subfigure{
      \includegraphics[scale=1]{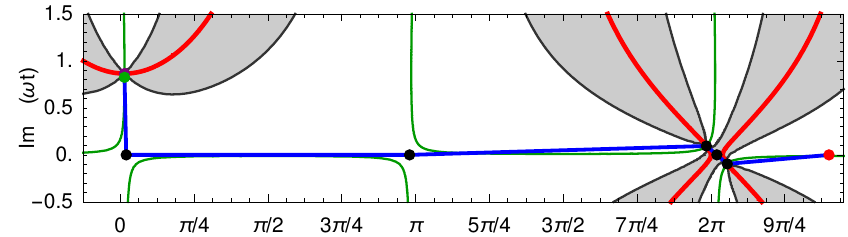}  
      \label{fig:path-chooser-examples-c}
    }
    \\
    (d) $\mathbf p=(\figurethirteendpo,0,\figurethirteendpp)$ \vspace{-3mm} \\
    \subfigure{
      \includegraphics[scale=1]{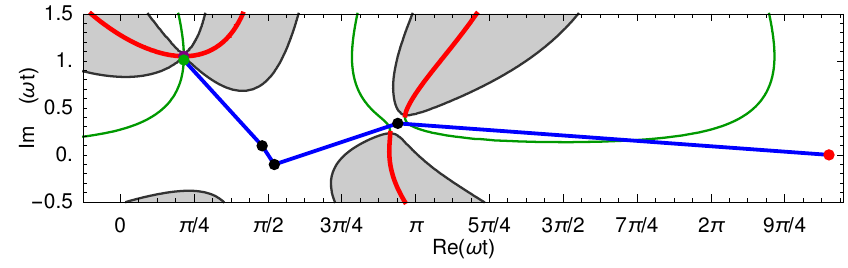}  
      \label{fig:path-chooser-examples-d}
    }
  \end{tabular}
  \caption{
       Sample integration contours produced by the $\tca$ choosing algorithm described in the text. Most momenta are straightforward \subref{fig:path-chooser-examples-a}, but near-recollision momenta, like the one shown in \reffig{fig:complex-position-branch-cuts}, do require careful handling, as shown in \subref{fig:path-chooser-examples-b}. The algorithm correctly handles soft recollisions, \subref{fig:path-chooser-examples-c}, as well as higher momenta with harder recollisions \subref{fig:path-chooser-examples-d}.
  }
  \label{fig:path-chooser-examples}
\end{figure}

We display in \reffig{fig:path-chooser-examples} some sample integration contours produced with this criterion. In general, the navigation is straightforward, and there are never any problems when $p_z<0$ or $p_\perp$ is sizeable, in which case contour looks as in \reffig{fig:path-chooser-examples-a}. For near-recolliding electrons, on the other hand, the branch cuts do require more careful navigation, as we have seen.

In general, it suffices to take, in order of increasing $\Re(\tca)$, those closest-approach times which (i) occur after ionization, (ii) have reasonably bounded imaginary part, and (iii) have positive kinetic energy. To this we add two exceptions: the first inward turning point, in $-\pi/2<\omega t <\pi/2$, which helps avoid crossing regions where $\Re(\vbv(t)^2)<0$ where this is not necessary; and the first closest-approach time, in $\pi/2<\omega t <3\pi/2$, when it can be consistently identified, which can sometimes have $\Re\left(\vbv(\tca)^2\right)<0$ at extremely large momenta, but is nevertheless the correct choice in those cases.

Thus, a concrete set of rules for choosing a contour is taking those $\tca$s for which
\begin{enumerate}[itemsep=-0mm]
\item[] $\Re(\omega \tca)>\omega\tn +\pi/5$ \textbf{and}
\item[] $-\frac13 \tauT<\Im(\tca)\leq \Im(\tk)$ \textbf{and}
\item[] $\Re\left(\vbv(\tca)^2\right) > -u$,
\item[\textbf{or}\hspace{3pt}] $-\pi/2 < \Re(\omega\tca) < \pi/2$ \textbf{and}
\item[] $0\leq\Im(\tca)<\tauT$,
\item[\textbf{or}\hspace{3pt}] $\pi/2 < \Re(\omega\tca) < 3\pi/2$ \textbf{and}
\item[] $\Im(\tca)>0$,
\end{enumerate}
where $u$ is an adjustable numerical precision, for additional flexibility, set by default to $\SI{e-8}{\au}$ These rules are relatively heuristic and they have a certain amount of leeway around them, but they work well  over the relevant region of photoelectron momentum to produce correct integration path choices. 


Finally, we note that in certain very specific cases, close to a soft recollision, the integration path is topologically correct but may pass near a singularity of the Coulomb kernel; future work aims at avoiding this problem.

\section{Results}
\label{sec:results}
We are now in a position to produce photoelectron spectra. The yield is given by the amplitude in \eqref{ARM-final-result}, taking the integration of \eqref{final-correction-factor} to be over the $\tca$ chosen using the hopping algorithm described above. The correctness of the integration contour is reflected by the lack of discontinuous changes in the final integrand, which are easily detected by numerical integration routines when they are present.

\begin{figure}[b]
  \begin{tabular}{c}
    \includegraphics[width=0.8\columnwidth]{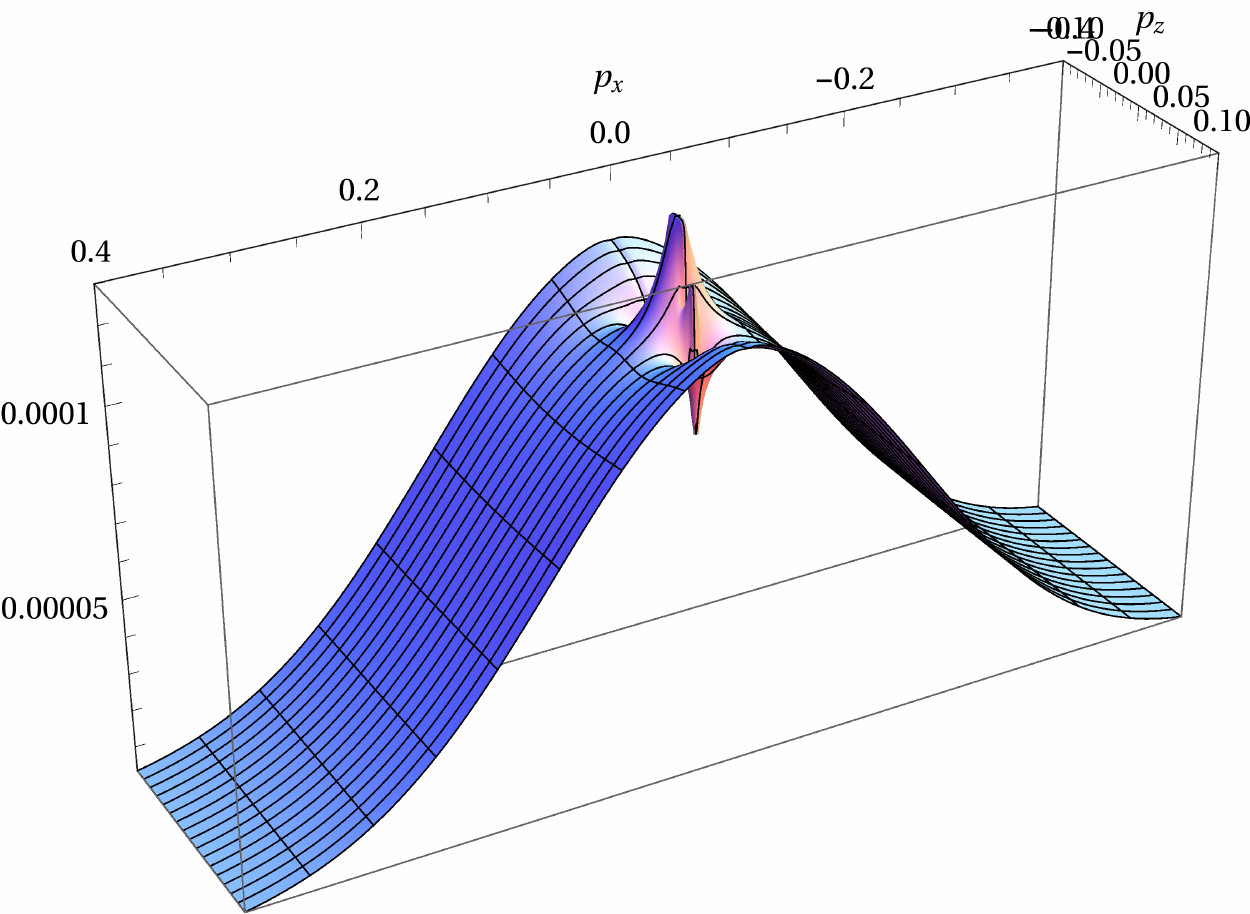}
  \end{tabular}
  \caption{
  Incoherent addition of the sub-cycle ionization yields, $\tfrac12\left(\left|a(p_x,0,p_z)\right|^2+\left|a(p_x,0,-p_z)\right|^2\right)$, ignoring the shape factor $R(\vbp)$, for long wavelength fields ($\lambda=\SI{\figurefourteenwavelength}{\micro\meter}$, with $\gamma=\figurefourteengamma$). The Coulomb-correction has been integrated over 2.75 laser periods. 
  Recent experimental results at this wavelength~\cite{pullen_kinematically_2014} show a similar (if more pronounced) peak structure localized at small longitudinal momentum.
  }
  \label{fig:po-pp-spectrum}
  \label{fig:po-pp-spectrum-comparison}
\end{figure}

In general, the inclusion of the Coulomb correction factor \eqref{final-correction-factor} has the main effect of increasing the electron yield by two or three orders of magnitude compared to the bare SFA yield, which is a well known consequence of the Coulomb interaction \cite{CCSFA_initial_short,TCSFA_sub_barrier,ARM_initial}, and is primarily due to the interactions inside the classically-forbidden region.

This enhancement is particularly strong, and has a particularly strong variation, near the classical soft recollisions $\pzsr$. As discussed above, these are accompanied by an extreme enhancement for momenta on one side of $\pzsr$ and a deep and narrow dip on the other. The presence of multiple saddle points in close proximity, which marks the increased time the electron spends near the core, implies the Coulomb correction must be large, but its precise contribution depends sensitively on the details of the topological transition at the soft recollision.

In the neighbourhood of such a soft recollision, the dominating feature is a sharp ridge for very small transverse momenta, as shown in \reffig{fig:po-pp-spectrum}, which terminates at $\pzsr$. This ridge feature is similar to the cusps detected at low transverse momenta in careful measurements of photoionization spectra at long wavelengths \cite{pullen_kinematically_2014}, which rise suddenly from a gaussian background as $p_\perp$ varies, though the measured spectra have sharper cusps.

On the other side of the soft recollisions, i.e. for $p_z>\pzsr$, this behaviour reverses, and the enhancement turns into a strong suppression of ionization. This suppression is evident in \reffig{fig:po-pp-spectrum}, which contains two transitions closely spaced at $\pzsr{}= \SI{ \figurefourteenfirstpztransition }{\au}$ and $\pzsr{}=\SI{\figurefourteenthirdpztransition}{\au}$; the interactions between them create complex structures in their immediate vicinity. These fine structures will most likely be washed out in any experimental situation, but the spike would be expected to remain.

\begin{figure}[b]
  \begin{tabular}{cl}
    \includegraphics[scale=1]{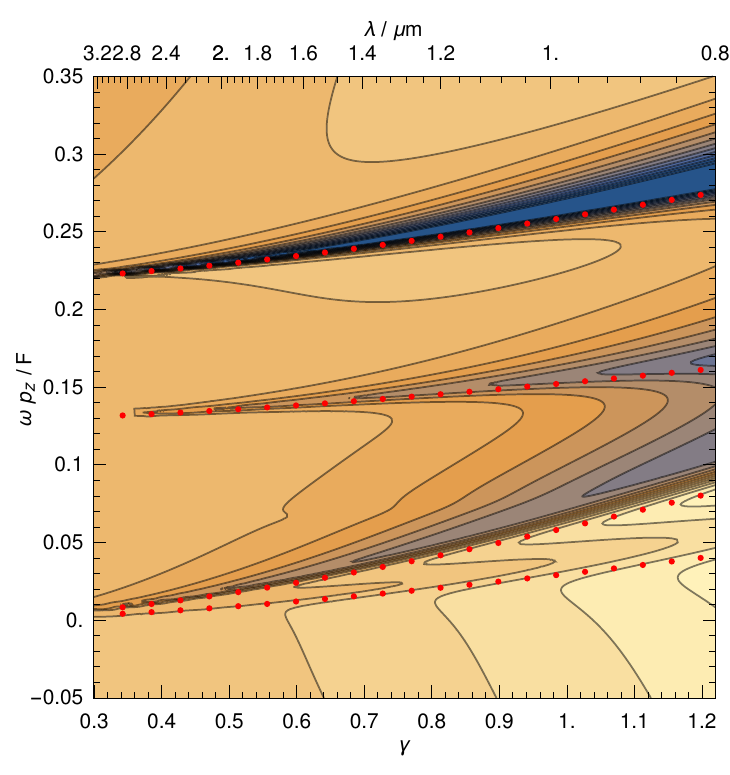}
    &
    \raisebox{8mm}{\includegraphics[scale=1]{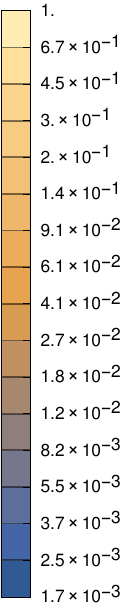}}
  \end{tabular}
  \caption{
  Variation of the on-axis ionization amplitude $|a(\vbp)|^2$, in an arbitrary logarithmic scale, as a function of the wavelength and the corresponding Keldysh parameter $\gamma$. The sudden drops in amplitude of \reffig{fig:po-pp-spectrum} shift along the momentum axis with a scaling that closely matches the classical soft-recollision trajectories, shown as red dots. 
  Here the transverse momentum $p_\perp$ has been chosen so that the transverse coordinate of the classical trajectory has a small but positive value, $x=\figurefifteenpomult{} \tfrac{1}{\kappa}$, at the first soft recollision at $\omega t\approx 2\pi$, to avoid the hard singularity of the Coulomb kernel.
  \vspace{-5mm}
  }
  \label{fig:scarred-spectrum}
\end{figure}

The rapid change from enhancement to suppression of the ionization signal at the topological transitions of \reffig{fig:branch-cut-topology-change} can be used to show their direct link with the classical soft recollisions of Section~\ref{sec:classical-transitions}, by comparing their respective wavelength scaling. To this end we plot in \reffig{fig:scarred-spectrum} the variation of the on-axis ionization yield -- at a small but reasonable transverse momentum -- with the laser wavelength. The resulting ridges in the ionization yield closely follow the scaling of the classical soft-recollision momenta of \reffig{fig:scaling}, shown as red dots, which shows a strong link between both structures. Similarly, the opposite scaling of the two types of low-energy structure is amenable to experimental testing.

\vspace{5mm}
\section{Conclusions}
\label{sec:conclusions}

We have seen how the analytical $R$-matrix theory gives rise, from first principles, to a quantum-orbit picture of strong field ionization which contains the Coulomb interaction with the ion as an added action in the phase. In this picture the electron trajectory starts at the ion with a real position, and as it travels through the classically forbidden region it acquires an imaginary component.

This imaginary component is not generally a problem, but it can dominate the trajectory near recollisions, in which case the Coulomb interaction can develop a discontinuous change in sign. This marks a branch cut which must be carefully handled, by treating the Coulomb interaction $U(\rcl(t))$ as an analytic function on the complex time plane, with branch cuts inherited from the spatial $U(\vbr)=-1/\sqrt{\vbr^2}$, which the temporal integration contour must avoid.

The key tool for avoiding these branch cuts in the complex time plane are the times of closest approach, which satisfy the complex equation $\rcl(\tca)\cdot\vbv(\tca)=0$, and which are always present in the middle of each branch-cut gate. These times of closest approach have a rich geometry of their own, in the complex quantum domain as well as in the real-valued simple-man's model. Moreover, the soft recollisions responsible for the Low Energy Structures are embedded in this geometry and arise naturally within both formalisms.

Using the times of closest approach, we have developed a consistent algorithm which enables the programmatic choice of a correct integration contour for all momenta, despite topological changes which can be quite brusque as the momentum crosses a soft recollision.

In particular, this formalism incorporates in a natural way the known series of trajectories which gives rise to the Low-Energy Structures, but it also predicts a second series of trajectories at much lower momentum. These trajectories do not appear in theories which neglect the tunnel entrance, which explains why they have so far been overlooked, but they should contribute to the recently discovered Zero-Energy Structures~\cite{ZES_paper,pullen_kinematically_2014,dura_ionization_2013}.

More broadly speaking, our work serves as a roadmap for the difficulties that must arise in a full first-principles semiclassical theory of ionization, and how they might be resolved. ARM theory, in its current form, uses only the laser-driven trajectory and does not incorporate the effect of the ion's Coulomb potential on the trajectory itself. However, it is also clear that any first-principles attempt to include the Coulomb interaction at that level will be subject to the same imaginary-position properties as ARM theory, which makes the calculation of the full trajectory a nontrivial problem. In particular, such a theory will be subject to the same temporal branch cuts as ARM theory, without the benefit of the easily-found ARM closest-approach times to help navigate them.

In addition, by identifying mechanisms which should contribute to the observed Zero Energy Structures, we are able to propose experiments which can increase its energy scale and bring its details within reach of current experimental capabilities. In particular, the energy scaling of odd-order soft-recollision trajectories as $\gamma^2 I_p$ suggests that experiments with harder targets, such as $\mathrm{He}^+$, should help elucidate the origin of these structures.

Furthermore, this trajectory-based explanation opens the door to testing via schemes which directly alter the shape of the trajectory. This includes variation of the pulse width, which is known to be a critical variable~\cite{Rost_JPhysB} along with the carrier-envelope phase, but it opens the door to testing using elliptical pulses or small amounts of second or third harmonics to shift the positions of the trajectories responsible for the structures.

This work thus provides a toolbox of approaches with which to understand the semiclassical dynamics of low-energy photoelectrons in the tunnelling regime, which should inform other low-energy ionization phenomena \cite{arbo_fan_shaped_interference,rudenko_fan_shaped_interference,off_axis_LES} as well as streaking-based experiments \cite{streaking_soft_recollisions} in that regime.

\begin{acknowledgments}
We thank L. Torlina, O. Smirnova, J. Marangos and F. Morales for helpful discussions. EP gratefully acknowledges funding from CONACYT and Imperial College London.
\end{acknowledgments}

\bibliographystyle{arthur} 
\bibliography{references}{}


\end{document}